\documentclass{aa}

\usepackage{graphicx} 
\usepackage{subfigure}
\usepackage{caption}
\usepackage{txfonts}
\usepackage{siunitx}
\usepackage{xspace}
\usepackage[switch]{lineno}
\usepackage[colorlinks,allcolors=blue]{hyperref}
\usepackage[dvipsnames]{xcolor}

\newcommand{\hess}{H.E.S.S.\xspace}
\newcommand{\gam}{$\gamma$\xspace}
\newcommand{\hwidth}{\text{Hillas-width}\xspace}
\newcommand{\hlength}{\text{Hillas-length}\xspace}
\newcommand{\hlogdens}{\text{Hillas-logdensity}\xspace}
\newcommand{\hlocdist}{\text{Hillas-local-distance}\xspace}
\newcommand{\hlols}{\text{Hillas-length-over-logsize}\xspace}
\newcommand{\hskew}{\text{Hillas-skewness}\xspace}
\newcommand{\hkurt}{\text{Hillas-kurtosis}\xspace}
\newcommand{\size}{\text{Hillas-image-amplitude}\xspace}
\newcommand{\npix}{\text{Hillas-npix}\xspace}
\newcommand{\npixI}{\text{npix$_{>10}$}\xspace}
\newcommand{\npixII}{\text{npix$_{>30}$}\xspace}
\newcommand{\ftI}{\text{fraction-top-1}\xspace}
\newcommand{\ftII}{\text{fraction-top-1+2}\xspace}
\newcommand{\kinkx}{\text{kink-major}\xspace}
\newcommand{\kinky}{\text{kink-minor}\xspace}
\newcommand{\kinkxgof}{\text{kink-major-GOF}\xspace}
\newcommand{\kinkygof}{\text{kink-minor-GOF}\xspace}
\newcommand{\roughI}{\text{roughness}\xspace}
\newcommand{\roughII}{\text{spatial-auto-correlation}\xspace}
\newcommand{\pg}{\text{profile-gradient}\xspace}
\newcommand{\pgI}{\text{profile-gradient}$_\mathrm{head}$\xspace}
\newcommand{\pgII}{\text{profile-gradient}$_\mathrm{tail}$\xspace}
\newcommand{\pggof}{\text{profile-gradient-GOF}\xspace}
\newcommand{\stda}{\text{std-over-mean}\xspace}
\newcommand{\distI}{\text{distance-top-1}\xspace}
\newcommand{\distII}{\text{distance-top-2}\xspace}
\newcommand{\distIII}{\text{distance-top-1-2}\xspace}
\newcommand{\tgx}{\text{time-gradient-major}\xspace}
\newcommand{\tgxgof}{\text{time-gradient-major-GOF}\xspace}
\newcommand{\tgy}{\text{time-gradient-minor}\xspace}
\newcommand{\tgygof}{\text{time-gradient-minor-GOF}\xspace}

\newcommand{\wO}{\text{width}$_\mathrm{head}$\xspace}
\newcommand{\wI}{\text{width}$_\mathrm{center}$\xspace}
\newcommand{\wII}{\text{width}$_\mathrm{tail}$\xspace}
\newcommand{\xO}{\text{x}$_\mathrm{head}$\xspace}
\newcommand{\xI}{\text{x}$_\mathrm{center}$\xspace}
\newcommand{\xII}{\text{x}$_\mathrm{tail}$\xspace}
\newcommand{\yO}{\text{y}$_\mathrm{head}$\xspace}
\newcommand{\yI}{\text{y}$_\mathrm{center}$\xspace}
\newcommand{\yII}{\text{y}$_\mathrm{tail}$\xspace}
\newcommand{\chargeasym}{\text{charge-asymmetry}\xspace}

\makeatletter
\renewcommand*\aa@pageof{, page \thepage{} of \pageref*{LastPage}}
\makeatother

\title{Improvements to monoscopic analysis for imaging atmospheric Cherenkov telescopes: Application to H.E.S.S.}

\titlerunning{Improvements to monoscopic analyses of IACTs}

\author{
  Tim Unbehaun\inst{\ref{ECAP}} \and
  Rodrigo Guedes Lang\inst{\ref{ECAP}} \and
  Anita Deka Baruah\inst{\ref{MPIK},\ref{IISER}} \and
  Prajath Bedur Ramesh\inst{\ref{MPIK},\ref{IISER}} \and
  Jelena Celic\inst{\ref{ECAP}} \and \\
  Lars Mohrmann\inst{\ref{MPIK}} \and
  Simon Steinmassl\inst{\ref{MPIK}} \and
  Laura Olivera-Nieto\inst{\ref{MPIK}} \and
  Jim Hinton\inst{\ref{MPIK}} \and
  Stefan Funk\inst{\ref{ECAP}}
}
\institute{
Friedrich-Alexander-Universit\"at Erlangen-N\"urnberg, Erlangen Centre for Astroparticle Physics, Nikolaus-Fiebiger-Str. 2, 91058 Erlangen, Germany\\ \email{tim.unbehaun@fau.de,rodrigo.lang@fau.de} \label{ECAP} \and
Max-Planck-Institut f\"ur Kernphysik, Saupfercheckweg 1, 69117 Heidelberg, Germany\\ \email{lars.mohrmann@mpi-hd.mpg.de} \label{MPIK} \and
Indian Institute of Science Education and Research (IISER) Tirupati, Tirupati 517507, Andhra Pradesh, India \label{IISER}
}

\date{January 15, 2025}

\abstract{
Imaging atmospheric Cherenkov telescopes (IACTs) detect \gam rays by measuring the Cherenkov light emitted by secondary particles in the air shower when the \gam rays hit the atmosphere of the Earth.
Given usual distances between telescopes in IACT arrays, at low energies ($\lesssim$ 100 GeV), the limited amount of Cherenkov light produced typically implies that the event is registered by one IACT only.
Such events are called monoscopic events, and their analysis is particularly difficult.
Challenges include the reconstruction of the event's arrival direction, energy, and the rejection of background events due to charged cosmic rays. 
Here, we present a set of improvements, including a machine-learning algorithm to determine the correct orientation of the image in the camera frame, an intensity-dependent selection cut that ensures optimal performance across all energies, and a collection of new image parameters.
To quantify these improvements, we make use of simulations and data from the 28-m diameter central telescope of the \hess IACT array.
Knowing the correct image orientation, which corresponds to the arrival direction of the photon in the camera frame, is especially important for the angular reconstruction, which could be improved in resolution by 57\% at 100\,GeV.
The event selection cut, which now depends on the total measured intensity of the events, leads to a reduction of the low-energy threshold for source analyses by $\sim$50\%.
The new image parameters characterize the intensity and time distribution within the recorded images and complement the traditionally used Hillas parameters in the machine learning algorithms.
We evaluate their importance to the algorithms in a systematic approach and carefully evaluate associated systematic uncertainties.
We find that including subsets of the new variables in machine-learning algorithms improves the reconstruction and background rejection, resulting in a sensitivity improved by 41\% at the low-energy threshold.
Finally, we apply the new analysis to data from the Crab Nebula and estimate systematic uncertainties introduced by the new method.
}

\begin{document}

\maketitle

\section{Introduction}
Since the success of the first generation of ground-based \gam-ray telescopes \citep{Weekes1989},  significant advances have been made in both hardware and software development~\citep{funk2015ground}. 
The basic detection principle has remained the same: Cherenkov light, emitted by charged secondary particles in \gam-ray induced air showers, is collected by imaging atmospheric Cherenkov telescopes (IACTs).
Modern arrays, such as VERITAS~\citep{weekes2002veritas}, MAGIC~\citep{MAGIC-2016}, and H.E.S.S.~\citep{hess-2006}, employ stereoscopic event reconstruction techniques, where one shower is observed by at least two telescopes.
This stereoscopic approach improves the reconstruction accuracy of the properties of the primary particle and improves the rejection of background events arising from interactions of charged cosmic rays~\citep{hegra1999performance}. 

However, not all \gam-ray events are captured by multiple telescopes, which makes monoscopic (or mono) event reconstruction still crucial.
This is particularly relevant for the \hess array, where the large central telescope, with its 28\,m mirror diameter, operates in mono-trigger mode and primarily detects low-energy events ($\lesssim$ 100 GeV), often seen by this telescope alone.
The importance and methodology for reconstructing these mono events are discussed in \citet{Murach2015}.
Their approach utilizes the traditional ``Hillas parameters'' \citep{hillas1985}, which characterize the first four moments of the intensity distribution within the shower images, as input to machine learning algorithms.
Although other reconstruction techniques exist, such as template-based methods \citep{Parsons2014}, background rejection for mono events in \hess is currently performed exclusively using image parameters.
Using intensity information directly in machine learning models is an alternative \citep[see, e.g.,][]{Holch-2017, parsons2020background, Lyard_2020, miener2022performancemagictelescopesusing, Glombitza_2023}, but poses challenges due to the sensitivity of the models to discrepancies between simulations and recorded data \citep{shilon-2019, parsons2022investigations}.
Improving the reconstruction and background rejection for low-energy \gam rays can further lower the energy threshold and increase the sensitivity of our observations.

By including the pixel time information in the analysis of monoscopic data from the MAGIC telescope, \citet{magic-mono-2009} could already achieve notable improvements in point-source sensitivity.
The construction and usage of additional image parameters in reconstruction algorithms, as well as the optimization of the event selection cut, is discussed in \citet{phd_voigt_2014}.
A recent study by \citet{lst2024} even included information from waveforms recorded by individual camera pixels in the reconstruction algorithms.

In this work, we explore the potential of incorporating additional image parameters that capture more detailed features of the amplitude and time distribution in the shower image. 
A systematic approach is used to assess the importance of each parameter for the reconstruction and background rejection algorithms, leading to an optimized subset of parameters tailored to each application.
We also examine the systematic uncertainties arising from discrepancies between different observation conditions, both prior to and following the inclusion of these new image parameters.
Moreover, we introduce a machine-learning algorithm trained to determine the direction of the shower development in the camera frame, enabling the calculation of the new image parameters based on a more accurate image orientation.
Currently, the direction of the shower is determined solely using the \hskew (defined in Sect.~\ref{sec:hillas-var}), which is an unreliable parameter for both small and truncated images.
We also implement a selection cut optimized for different event sizes (size here and in what follows stands for the total image intensity), which outperforms the standard constant selection cut used in \citet{Murach2015}.
This optimization is crucial because the performance of the classification algorithm improves significantly with event size, and a constant cut does not fully exploit the potential across all sizes \citep{voigt2014threshold}.

The new image parameters are introduced in Sect.~\ref{sec:new-variables}, followed by the description of the machine learning model architecture for event reconstruction and classification in Sect.~\ref{sec:architecture}, and the optimization of the size-dependent selection cut in Sect.~\ref{sec:size-dependent-sel-cut}.
Finally, we assess the performance of the new reconstruction chain and compare it to the previous one in Sects.~\ref{sec:event-reco}-\ref{sec:sensitivity}, with an application to real Crab Nebula data presented in Sect.~\ref{sec:data}.

\section{Implementation}
This study is based on \gam-ray events simulated and observed using the central large telescope (CT5) of the \hess array, equipped with the FlashCam camera \citep{vanEldik2016, FC-Bi-2021}.
Air showers are simulated using the \texttt{CORSIKA} package \citep{heck1998corsika}, and the telescope response is simulated using \texttt{sim\_telarray} \citep{simtel-2008}. 
Simulated events follow a power-law distribution $\sim$\,$E^{-\gamma}$ with a spectral index of 1.8 to increase statistics at higher energies and are then re-weighted to a spectral index of 2.5.
After extracting the pixel amplitudes and times (i.e., integrated photoelectrons (p.e.) and peak signal times), noise pixels are set to zero by a cleaning procedure. 
A tail-cut image cleaning is used to classify core pixels, which have an amplitude above 14\,p.e and at least 2 neighbor pixels above 7\,p.e. Core and neighbor pixels are kept, while the time information is not considered during this step.  
Algorithms that use time information for image cleaning have been developed for \hess~\cite{phd_simon_2023} but are not yet adopted in the standard analysis pipeline. 
For this reason, we adhere to the standard tail-cut method to maintain consistency and minimize introducing correlated changes into the analysis pipeline. 
Future studies will explore potential enhancements to the image cleaning technique and the resulting improvements in source detection sensitivity.
Based on the remaining signal pixels, image parameters are computed, forming the basis for event reconstruction and classification.
All these steps (after simulation) are implemented in the \hess analysis program (HAP). 
For this study, we apply preselection cuts, requiring a total image amplitude above 80\,p.e. and a minimum of five pixels.
This minimum image size is relatively close to the trigger threshold of the telescope at 69\,p.e. \citep{FC-Bi-2021}, while maintaining a safe distance, to avoid systematic uncertainties due to mis-modeling of the camera response in this regime.
The distance from the image center of gravity to the center of the camera (so-called \hlocdist, see Sect.~\ref{sec:hillas-var}) is required to be smaller than 0.8\,m to reduce image truncation effects at the camera edges, which are between \SIrange{1.0}{1.2}{m}.

In the following subsections, we define the image parameters investigated in this study before presenting the reconstruction framework and architecture of the machine learning models.

\subsection{Definition of training variables}
\label{sec:new-variables}

The variables described in this section characterize the spatial and temporal distribution of the pixel amplitudes and times, either for the whole image or parts of the image.
To improve background rejection, they aim to be sensitive to differences between cosmic ray-induced (hereafter hadronic) background showers and \gam ray-induced air showers.
These differences are most prominently seen in the more irregular shower morphology for hadronic showers, caused by electromagnetic, muonic, and hadronic subshowers.
In Appx.~\ref{appx:new-var}, the distribution of each variable is shown for measured background data, together with proton and \gam-ray simulations.

Even though one would not necessarily expect perfect agreement between background data and proton simulations due to various factors, such as the primary cosmic-ray composition or atmospheric and instrument conditions, a good agreement between these distributions indicates that the simulated atmosphere, telescope, and night sky background (NSB) light match reality.
The consistency between the measured and simulated data for FlashCam in CT5 is discussed in detail by \citet{leuschner2023validating}.
Since the background events are characterized by their isotropic arrival directions, we compare their distributions to those of diffuse \gam-ray simulations\footnote{Simulations with isotropic arrival directions, as opposed to point-source simulations with only one arrival direction.}.
These are available with a realistic NSB value, based on the average observed Galactic NSB, and a value that is $\sim$\,1.65 times higher than this.
A difference between these sets of simulations indicates that the given variable is sensitive to the details of the NSB and therefore introduces systematic uncertainties, as the NSB values vary between observations but are typically kept at a fixed value for most simulations. 
Large differences between the hadronic and \gam-ray distributions imply strong separation power for this variable, although some variables are correlated, which reduces the gain in separation power when combining these variables.

The camera image reflects the development of the shower in the atmosphere and is typically parameterized as an ellipse. 
The longitudinal development is encoded along the major axis of the image, the head reflecting the top part of the shower where the first interaction takes place, while the tail represents the bottom part of the shower where many shower particles contribute to the emission. 
Since the charged shower particles travel at a slightly higher speed compared to the Cherenkov photons in the atmosphere, the arrival times do not necessarily follow the sequence head-center-tail.
They rather depend on the impact distance, the distance between the shower's impact point on the ground and the telescope's position, for geometrical reasons.

For the computation of the variables, we consider the position of pixel $i$ in the camera frame $(X_\mathrm{CAM,i}, Y_\mathrm{CAM,i})$, its amplitude $A_i$ (integrated charge in the readout window), and time $T_i$ (peak time of the charge time-evolution).
The image coordinates $(X_i,Y_i)$ then follow as a translation plus rotation
\begin{equation}
    \begin{bmatrix} X_i \\ Y_i \end{bmatrix} = \begin{bmatrix} \cos(\alpha) & \sin(\alpha) \\ -\sin(\alpha) & \cos(\alpha) \end{bmatrix} \begin{bmatrix} X_\mathrm{CAM,i} - {<}X_\mathrm{CAM}{>} \\ Y_\mathrm{CAM,i} - {<}Y_\mathrm{CAM}{>} \end{bmatrix} \, ,
\end{equation}
where $\alpha$ is the angle that minimizes the r.m.s. of the intensity distribution along the $Y$-direction and $({<}X_\mathrm{CAM}{>},{<}Y_\mathrm{CAM}{>})$ is the center of gravity of the image in the camera frame.
Consequently, the center of gravity lies at (0,0) in these coordinates, and the major axis of the ellipse is aligned with the $X$ axis.

\subsubsection{Hillas parameter-based variables}
\label{sec:hillas-var}
The first set of variables was introduced by \citet{hillas1985} and characterizes the second, third, and fourth moments of the amplitude distribution.
These variables are widely referred to as ``Hillas parameters,'' and their calculation is described in Appendix A of \citet{reynolds1993survey}.

\paragraph{``\size''} is the total amplitude summed over all pixels $\sum_i A_i$.
It is mainly a function of the \gam-ray energy and impact distance on the ground.
Additionally, the atmospheric transparency and telescope efficiency affect the total light yield of the showers.
Therefore, the distribution of the \size (or size, cf. Fig.~\ref{fig:dist-size}) is very similar for different primary particle showers, in particular if they follow similar energy spectra. It cannot and should not be used as a separating variable.
However, since the gamma-hadron separation works better for images with a larger total amount of light in the camera, information about the size is very useful for separation and is implicitly available through correlations with other variables.

\paragraph{``\npix''} is the number of pixels with $A_i > 0$ that pass the aforementioned image cleaning.
Its distribution is shown in Fig.~\ref{fig:dist-npix}.
\npix is affected by the same shower properties as the \size, and additionally by the shower morphology and the image cleaning.

\paragraph{``\hlocdist''} is the distance from the image center of gravity to the center of the camera, defined as $\sqrt{{<}X_\mathrm{CAM}{>}^2 + {<}Y_\mathrm{CAM}{>}^2}$. 
This parameter is solely determined by the position of the shower in the camera's field of view. 
Although it does not contain any physical information beyond the position, it is of interest to algorithms, as images recorded near the edges of the camera may require a different handling due to truncation effects.

\paragraph{``\hwidth''} is defined as the r.m.s. along the $Y$-direction and can be computed as:
\begin{equation}
\label{eq:hwidth}
    \text{\hwidth} = \sqrt{ \frac{\sum_i Y_i^2 \cdot A_i}{\sum_i A_i} }
\end{equation}
with the distribution shown in Fig.~\ref{fig:dist-width}.
The characteristic dip located at $\sim$\,\SI{2}{cm}, slightly below half of the camera pixel width (\SI{5}{cm}), occurs because images with only 5, 6, or 7 pixels have a sharp cutoff in their width distribution at this value.
The vast majority of possible configurations for 5, 6, or 7 connected pixels in the hexagonal camera frame have their largest extent along one of the three symmetry axes, resulting in preferred directions of $\alpha$.
Therefore, most images with low pixel numbers have only two ``rows'' of pixels along the $Y$-direction, with a maximal r.m.s. of half the pixel width.
For images with more pixels, the distribution shifts toward larger widths and becomes more and more continuous.

\paragraph{``\hlength''} is defined analogously as the r.m.s. along the $X$-direction:
\begin{equation}
    \hlength = \sqrt{ \frac{\sum_i X_i^2 \cdot A_i}{\sum_i A_i} } \, 
\end{equation}
with the distributions shown in Fig.~\ref{fig:dist-length}.
The length and, especially, the width distributions show differences between hadrons and \gam rays, with the latter being more abundant at smaller values, as expected for more compact and regular showers.

The third and fourth moments along the major axis can be computed as:
\begin{equation}
    K_m = \frac{\sum_i X_i^m \cdot A_i}{\sum_i A_i} \cdot \frac{1}{\hlength^m} \, ,
\end{equation}
where $K_m$ is the m'th moment.

\paragraph{``\hskew''} is the third moment $K_3$ with the distribution shown in Fig.~\ref{fig:dist-skew}.
There are minor differences in the distributions, as the more irregular proton showers feature larger absolute values of skewness, indicating a broader distribution.

\paragraph{``\hkurt''} is the fourth moment $K_4$ with the distribution shown in Fig.~\ref{fig:dist-kurt}.
The more irregular proton showers with their broader intensity distribution feature lower values of kurtosis.

\paragraph{``\hlogdens''} is a composite variable that includes the \size and is defined as:
\begin{equation}
    \hlogdens = \log_{10}\left(\frac{\size}{\hwidth \cdot \hlength}\right).
\end{equation}
The \size does not enter the separation training directly, to avoid a strong dependence of the algorithm on the simulated spectrum, but is implicitly available through the composite variables.
The distribution is shown in Fig.~\ref{fig:dist-logdens}, where \gam-ray showers have a tail toward higher densities, while hadronic showers tend toward lower densities. 
This matches the \npix distributions, where the latter tend toward higher pixel numbers.

\paragraph{``\hlols''} is also a composite variable and defined as:
\begin{equation}
    \hlols = \frac{\hlength}{\log_{10}(\size)}.
\end{equation}
The distribution of the \hlols is shown in Fig.~\ref{fig:dist-lols} and is very similar to that of the \hlength.

\subsubsection{Sector-based variables}
The following image parameters (also those introduced in Sects.~\ref{sec:continuous-var} and~\ref{sec:pix-variables}) are part of a set of novel variables that have not yet been utilized in any previous analyses. 
These parameters are specifically designed to enhance both the reconstruction accuracy and background rejection in monoscopic IACT images.

\begin{figure}[h]
    \centering
    \includegraphics[width=.98\linewidth]{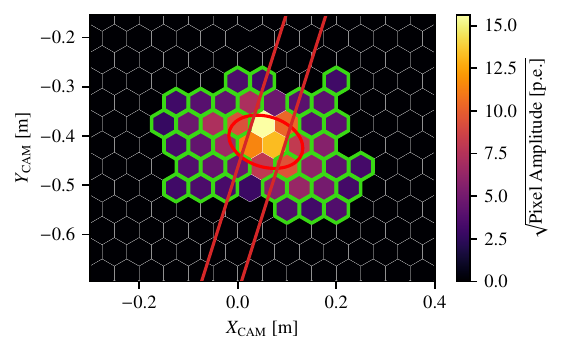}
    \caption{
        Simulated \gam-ray event in the camera frame overlaid with a red ellipse illustrating the Hillas parametrization (width and length), and the lines separating the sectors.
        The pixels marked in green belong to either the head or the tail sector, while pixels that are not marked belong to the central sector.
     }
    \label{fig:example-event-sectors}
\end{figure}

The idea behind the sector-based variables is to increase the sensitivity to asymmetries in the image that arise from the asymmetric development of the shower in the atmosphere, such as the head-tail asymmetry, or potential asymmetries in the hadronic shower development.
For this, we divide the image into three sectors along the major axis: the head, center, and tail sectors, and compute moments of the charge distribution in these sectors separately.
When $L=\hlength$, all pixels with $X_i < -0.5\cdot L$ belong to the first sector, pixels with $|X_i| < 0.5\cdot L$ belong to the center sector, and pixels with $X_i > 0.5\cdot L$ belong to the third sector (cf. Fig.~\ref{fig:example-event-sectors}).
Due to the random image orientation in the camera, the assignment of sectors one and three to the head and tail sectors occurs later, after the correct image orientation is estimated (see Sect.~\ref{sec:flip_algo}). 

For each of the sectors, we compute ${<}X{>}$ and ${<}Y{>}$, resulting in the variables ``\xO,'' ``\xI,'' and ``\xII,'' and ``\yO,'' ``\yI,'' and ``\yII,'' with the distributions shown in Figs.~\ref{fig:dist-xO}$-$\ref{fig:dist-yII}.

Additionally, we compute the width according to Eq.~\ref{eq:hwidth} for each of the sectors, resulting in the variables ``\wO,'' ``\wI,'' and ``\wII'' (cf. Figs.~\ref{fig:dist-wO}$-$\ref{fig:dist-wII}).

The peaks at zero for the distributions of the center sectors are caused by low-size events where the central sector has zero pixels.
This corresponds to a division of the event into only two sectors.
Variations of ${<}Y{>}$ from zero indicate a curvature of the image along the major axis, which is typical for high-impact distance muons where only part of the characteristic muon ring is visible.
However, this feature may be better captured by the continuous ``\kinkx'' variable described in Sect.~\ref{sec:continuous-var}.
While \wI is strongly correlated with the \hwidth, a comparison of the head and tail widths allows the characterization of a gradient in the width along the major axis.
This is then correlated with the continuous variable ``\kinky'' (cf. Sect.~\ref{sec:continuous-var}).
The ${<}X{>}$ values for the head and tail sectors of the \gam-ray showers are slightly closer to zero, which is connected to the larger kurtosis.
Asymmetries between the head and tail values for one event are reflected in the skewness. 

Finally, one can also compute the ``\chargeasym'' as:
\begin{equation}
    \chargeasym = \frac{\sum_i A_{i,\mathrm{head}} - \sum_i A_{i,\mathrm{tail}}}{\sum_i A_i}\, ,
\end{equation}
which is also correlated with the \hskew and has a very similar distribution for background data and \gam-ray simulations (cf. Fig.~\ref{fig:dist-chargeasym}).

\subsubsection{Continuous variables}
\label{sec:continuous-var}
In the following, we aim to characterize additional features of the event images that the variables introduced up to here are not sensitive to.
As the sector variables are not optimal, especially for low-size events, we no longer separate the pixels into three sectors but rather use all pixels to compute gradients and other correlations.
The first set of variables is based on the gradient $m$ between two pixel attributes $x$ and $y$, which is computed via linear regression: 
\begin{equation}
\label{eq:m}
    m = \frac{\sum_i (x_i - {<}x{>})\cdot (y_i - {<}y{>})\cdot w_i}{\sum_i w_i \cdot \mathrm{Var}(x)} \, ,
\end{equation}
where ${<}x{>}$, ${<}y{>}$ are the weighted averages of $x$ and $y$, respectively, using the weights $w$.
Var$(x)$ is the weighted variance computed as follows:
\begin{equation}
    \mathrm{Var}(x) = \frac{\sum_i (x_i-{<}x{>})^2 \cdot w_i}{\sum_i w_i}\, ,
\end{equation}
again with ${<}x{>}$ being the weighted average of $x$.
The y-value $t$ at $x=0$ is computed as:
\begin{equation}
\label{eq:t}
    t = {<}y{>} - m\cdot {<}x{>}
\end{equation}
and the goodness-of-fit (GOF) or weighted squared deviation $\chi^2$ from the linear curve:
\begin{equation}
\label{eq:chi2}
    \chi^2 = \frac{\sum_i (y_i - x_i\cdot m + t)^2 \cdot w_i}{\sum_i w_i}\,.
\end{equation}

\begin{figure}[h]
    \centering
    \includegraphics[width=.98\linewidth]{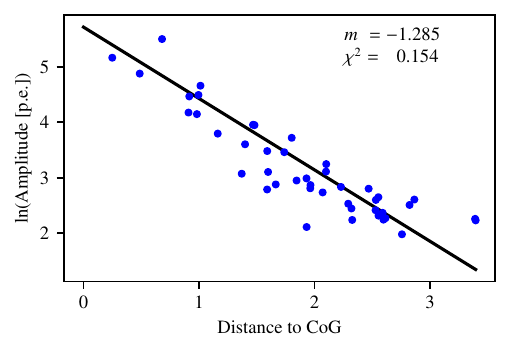}
    \caption{
        Profile-gradient for a simulated \gam-ray event. The distance to the CoG is computed according to Eq.~\ref{eq:r} and is therefore unitless.
     }
    \label{fig:example-pg}
\end{figure}

\paragraph{``Profile-gradient''} The first example is the \pg, where we compute the slope with which the logarithm of the pixel amplitudes $A_i$ changes with distance to the center of gravity (CoG) of the image (cf. Fig.~\ref{fig:example-pg}).
The distance to the CoG $r$ is computed as:
\begin{equation}
\label{eq:r}
    r = \sqrt{\left(\frac{X}{\hlength}\right)^2 + \left(\frac{Y}{\hwidth}\right)^2}\,.
\end{equation}
The \pg follows as slope $m$ of Eq.~\ref{eq:m} with $x=r$, $y=\ln(A\,[\mathrm{p.e.}])$, and $w=A$. 
Using the same variables and the resulting $m$ and $t$, one can also compute the spread around the \pg as $\chi^2$ from Eq.~\ref{eq:chi2}, resulting in the ``\pggof.''

Compared to \hkurt, \pg is sensitive not only to the charge distribution along the major axis but also along the minor axis.
The distribution of \pg in Fig.~\ref{fig:dist-pg} shows that \gam-ray showers usually have larger negative gradients, typical for more compact showers. 
Very few, dominantly hadronic, showers have positive gradients, which are characteristic of muon rings where the intensity increases toward the edges.
The distribution of the \pggof, shown in Fig.~\ref{fig:dist-pggof}, also indicates that proton showers are more irregular and therefore have a larger spread around the linear gradient.
One can also compute the \pg for pixels with $X_i > 0$, resulting in ``\pgI,'' and for pixels with $X_i < 0$, resulting in ``\pgII.'' 
For asymmetric events, these values can be quite different.

\paragraph{``Kink-major''} is the gradient $m$ from Eq.~\ref{eq:m}, with $x=|X|$, $y=Y$, and $w=A$, and is sensitive to the curvature of the ellipse along the major axis.
When muons are recorded at large impact distances, only part of the muon ring is visible in the camera, which might be confused with a \gam-ray ellipse, except for its curvature.
Fig.~\ref{fig:dist-kinkx} shows that \gam-ray events tend to have values of \kinkx closer to zero, while some hadronic events show larger absolute values for \kinkx.
The calculation of the ``\kinkxgof'' as $\chi^2$ of Eq.~\ref{eq:chi2} is identical to the definition of the \hwidth in the case of \kinkx$=0$.
For values of \kinkx$\neq 0$, the \kinkxgof is smaller than the \hwidth due to one more degree of freedom.
To further reduce the correlation with \hwidth and to make the distribution more similar for events of different sizes, we divide the \kinkxgof additionally by $\npix -2$, as $m$ and $t$ represent two additional degrees of freedom.
With this computation, the distribution is smoothed and the characteristic dip present in the distribution of \hwidth is eliminated.

\paragraph{``Kink-minor''} is defined similarly as $m$ from Eq.~\ref{eq:m} with $x=Y^2$, $y=X$, and $w=A$, and represents a gradient in ${<}X{>}$ for increasing offsets to the major axis.
This is tightly connected to a change in the ellipse's width along the major axis. 
We also tested the variable resulting from $y=|Y|$ but found that the separation between \gam-ray and hadronic events is not as good, as the influence of low-intensity pixels far away from the major axis is reduced.
From Fig.~\ref{fig:dist-kinky}, one can see that hadronic showers tend to have values closer to zero, while the \gam-ray simulations show larger tendencies for more asymmetric images.
The sign of the \kinky is especially interesting, as shower images are expected to have a larger width in the tail of the shower facing the ground, so the sign of \kinky is useful in the angular reconstruction of the events.
The ``\kinkygof'' follows as $\chi^2 / (\npix -2)$ and is proportional to the \hlength.
Figure~\ref{fig:dist-kinkygof} shows that the distributions for \gam rays and background events are similar.

\paragraph{``Time-gradient-major''} 
In the variables described so far, the time information in the pixels is not yet used.
We compute the \tgx as $m$ from Eq.~\ref{eq:m} with $x=X$, $y=T$, and $w=A$, and the ``\tgy'' with $x=Y$ and $y=T$, and $w=A$.
The distributions for the two time gradients are shown in Figs.~\ref{fig:dist-tgx} and~\ref{fig:dist-tgy}.
While the \tgy is expected to be close to zero for all showers, the \tgx is expected to be mainly a function of the impact distance for geometrical reasons.
Therefore, the difference between the distributions of \gam-ray and hadronic showers is somewhat surprising.
As we will see in Sect.~\ref{sec:AD-test}, however, these variables do not pass our selection procedure.

The difference between the time spreads ``\tgxgof'' and ``\tgygof,'' shown in Figs.~\ref{fig:dist-tgxgof} and~\ref{fig:dist-tgygof}, is highly affected by different NSB rates.
While this is expected for the tail-cut image cleaning, which is neglecting the pixel's time information, we cannot use these variables without introducing large systematic uncertainties.

\paragraph{``Roughness''}
The next set of variables is meant to be sensitive to the regularity or smoothness of the shower.
One can compare a smoothed version of the image to the original by adding the differences for each pixel.
This is done in the \roughI variable, which is calculated as follows:
\begin{equation}
    \roughI = \sum_i \left(\frac{\sum_j (A_i - A_j)^2 \cdot w_{ij}}{\sum_j w_{ij}}\right) \frac{\npix}{\size^2}
\end{equation}
where $w_{ij}$ represents weights of a Gaussian kernel with $\sigma = 1\,\mathrm{pixel~width}$:
\begin{equation}
\label{eq:wij}
    w_{ij} = \exp\left( - \frac{(X_i - X_j)^2 + (Y_i - Y_j)^2}{2 \sigma^2}\right)\,.
\end{equation}
The \roughI variable adds up differences in the pixel amplitude between neighboring pixels, while the ``\stda'' variable adds differences to the mean from all shower pixels.

\paragraph{``Std-over-mean''} converges to the \roughI parameter in the limit of large kernel widths.
It is computed as:
\begin{equation}
    \stda = \frac{\sqrt{\frac{\sum_i (A_i - {<}A{>})^2}{\npix -1 }}}{{<}A{>}}\,.
\end{equation}
Therefore, the variables \roughI and \stda are correlated, but the strength of the correlation is different for \gam rays (0.88) and background events (0.79).
The distribution for both variables can be seen in Figs.~\ref{fig:dist-rough1} and \ref{fig:dist-stda}.
One can see that for both variables, the \gam-ray simulations tend toward larger values due to their more compact shape, leading to larger differences in the amplitudes of neighboring pixels.

\paragraph{``Spatial-auto-correlation''}
An attempt to characterize differences from the expected event morphology resulted in the computation of the \roughII (Moran's $I$) of the residuals after subtracting the expected pixel amplitude based on the \pg:
\begin{equation}
    I = \frac{N \cdot (N-1)}{W}\frac{\sum_i \sum_j (R_i - {<}R{>})(R_j - {<}R{>})\cdot w_{ij}}{\sum_i (R_i - {<}R{>})^2}\,,
\end{equation}
with $N = \npix$, $w_{ij}$ the weights from Eq.~\ref{eq:wij}, and $W = \sum_i \sum_j w_{ij}$.
$R_i$ is calculated from the previously determined \pg $m$ and the corresponding $t$ (cf. Eq.~\ref{eq:m} and \ref{eq:t}):
\begin{equation}
    R_i = \frac{A_i - \exp(r_i\cdot m + t)}{\sqrt{\exp(r_i\cdot m + t)}}
\end{equation}
and
\begin{equation}
    {<}R{>} = \frac{\sum_i R_i}{N}\,.
\end{equation}
In the absence of any correlation, the expected value for $I$ is $-1$, with more negative values indicating spatial anticorrelation and more positive values indicating spatial correlation.
The distributions of the \roughII shown in Fig.~\ref{fig:dist-rough2} are very similar between background and \gam rays, with background events slightly more abundant at larger negative values.

\subsubsection{Pixel Variables}
\label{sec:pix-variables}
To improve gamma-hadron separation, especially for low-size images, we also examine distributions focusing on the two brightest pixels.
From their amplitude values $A1$ and $A2$, we compute ``\ftI'' and ``\ftII'' as
\begin{equation}
    \ftI = \frac{A1}{\size}
\end{equation}
and 
\begin{equation}
    \ftII = \frac{A1+A2}{\size}\,,
\end{equation}
respectively.
If the machine learning algorithm has access to both values, the ratio between the two brightest pixels is also accessible.
The distributions in Figs.~\ref{fig:dist-ft1} and~\ref{fig:dist-ft2} show that \gam-ray showers have a larger charge concentration in the brightest pixel(s).

The other three pixel parameters concern the position of the brightest two pixels with respect to the CoG and to each other:
\begin{equation}
    \distI = \frac{\sqrt{(X1-{<}X{>})^2 + (Y1-{<}Y{>})^2}}{\hlength},
\end{equation}
and
\begin{equation}
    \distII = \frac{\sqrt{(X2-{<}X{>})^2 + (Y2-{<}Y{>})^2}}{\hlength},
\end{equation}
and
\begin{equation}
    \distIII = \frac{\sqrt{(X1-X2)^2 + (Y1-Y2)^2}}{\hlength}\,,
\end{equation}
with $(X1, Y1)$ being the coordinates of the brightest pixel, $(X2, Y2)$ the coordinates of the second brightest pixel, and $({<}X{>}, {<}Y{>})$ the coordinates of the CoG.

The distributions in Figs.~\ref{fig:dist-dist1}-\ref{fig:dist-dist12} show that for \gam-ray showers, it is more likely to find the brightest, but also the second brightest, pixel closer to the CoG, which is expected for more regular showers.
For the distance between the two brightest pixels, there is a plateau in the distributions around one \hlength, which is more concentrated for \gam-ray showers, while there is a larger spread for hadronic showers.

Finally, simply counting the number of pixels that exceed the thresholds of 10\,p.e. and 30\,p.e. results in the variables ``\npixI'' and ``\npixII,'' respectively.
Figures~\ref{fig:dist-n10} and~\ref{fig:dist-n30} show the distributions of these variables, where small \gam-ray images tend to have brighter pixels, while for larger images, hadronic showers have a more homogeneous light distribution.

\subsubsection{NSB sensitivity of the image parameters}
\label{sec:AD-test}
By introducing additional variables, one not only increases sensitivity to physical differences between the shower types, but also to systematic differences between data and simulations.
One possible source of systematic difference is the NSB level, which varies within the field of view (FoV) and between different observations, but is simulated at a constant averaged value.
To obtain a more quantitative estimate of the NSB sensitivity per variable, we use point-source simulations for nominal and high NSB values.
These simulations are more efficient than the diffuse simulations introduced earlier, which leads to a statistically more accurate estimate.
As for the diffuse simulations, the high NSB value is 1.65 times larger compared to the nominal NSB value and exceeds the NSB values of the vast majority of observations.

\begin{figure}
    \centering
    \includegraphics[width=.98\linewidth]{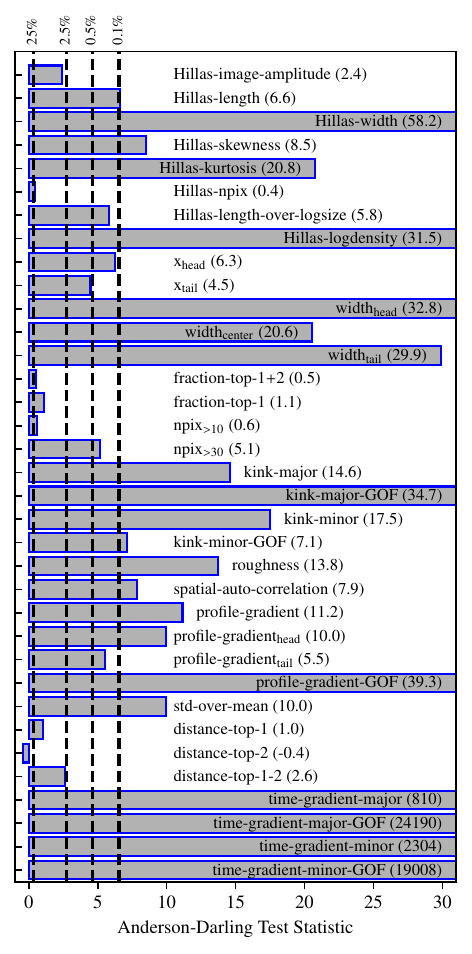}
    \caption{
        Anderson-Darling statistic for the distribution of nominal and high-NSB simulations. Larger values correspond to a more significant difference between the distributions. The vertical lines show four selected significance levels at 25\%, 2.5\%, 0.5\%, and 0.1\%. The numbers next to the variable names indicate the value of the Anderson-Darling statistic since some of the values extend beyond the plotted range.
     }
    \label{fig:anderson-stat}
\end{figure}

For each variable, we employ an Anderson-Darling test for k-samples, which tests the null hypothesis that the two distributions are drawn from the same population.
Figure~\ref{fig:anderson-stat} summarizes the results of these tests.

The danger of an NSB-sensitive distribution is that the separation training will perform differently on observations taken under different NSB conditions, introducing systematic uncertainties in the reconstructed \gam-ray rates.
Additionally, the performance might also change across the FoV introducing background artifacts, like gradients.

Compared to all other variables, the time variables show the most significant NSB dependence.
This is because the image cleaning completely ignores the pixel times and their clustering for shower pixels.
Consequently, higher NSB rates lead to a higher rate of noise pixels with larger amplitudes that survive the image cleaning, but which are randomly distributed in time.
These pixels can drastically change the time gradient, if the random pixel time is very different from the shower time.
Therefore, we do not use time-based variables for the separation training.

A time-based image-cleaning technique might solve this issue and would allow time variables to be used in the separation training.
The exact influence of the NSB rates on the effective area and thus the flux systematics will be evaluated in Sect.~\ref{sec:bkg-rejection}.

\subsection{Architecture of training algorithms}
\label{sec:architecture}

The event reconstruction used in this work is based on the current \hess monoscopic analysis procedure. 
The first step involves determining the so-called ``flip'' ($F$), which specifies whether the image center is to the left or right of the true source position in the camera frame.
The source is expected to be located along the major axis of the Hillas ellipse. 
However, for a monoscopic reconstruction, it is challenging to determine the direction along the axis. 

Next, the distance between the center of gravity of the ellipse and the source location, called ``disp'' ($\delta$), is reconstructed \citep{Hofmann_1999}. 
The reconstructed source position in the camera reference frame can then be obtained using the following equations:
\begin{equation}
\begin{cases}
    X_{\rm{CAM}}^{\rm{rec}} = X_{\rm{CAM}}^{\rm{COG}} +  F \cos(\alpha) \delta \\
    Y_{\rm{CAM}}^{\rm{rec}} = Y_{\rm{CAM}}^{\rm{COG}} +  F \sin(\alpha) \delta
\end{cases}
,
\end{equation}
where $F = \pm 1$, and $\alpha$ is the angle of the major axis of the ellipse with respect to the camera frame.

In the standard \hess analysis, the order in which flip and disp are determined is irrelevant. 
However, for this work, some of the new variables introduced depend on distinguishing the head and tail of the ellipse. 
Thus, the flip must be determined first and if it is positive, these variables change sign or are interchanged with their corresponding variables on the opposite side of the shower.
The flipping behavior for each variable is detailed in Tab.~\ref{tab:var_summary}.

Moreover, the flip-dependent image parameters are also sensitive to the position of the source in the camera frame, while the simulations are always performed for a point source at $X_{\rm{CAM}}=+0.5^\circ$.
This would normally lead to an asymmetry in the flip distribution, since the region left to the source position is larger than that to the right.
Therefore, we mirror half of the simulated events with respect to the $Y_{\rm{CAM}}$ axis ($X_{\rm{CAM}} \rightarrow -X_{\rm{CAM}}$), randomizing the distributions.
With that, we ensure a performance independent of the camera sector.

The subsequent steps involve energy regression and the estimation of the event's gammaness, that is, a quantity reflecting the probability that the primary particle was a \gam ray, compared to a hadronic particle. 
Also in these trainings, the flip determination and mirror operations described above are required beforehand due to the head- and tail-dependent variables being used. 
Aside from this, both reconstructions are independent of the other reconstruction steps.
Here, we evaluate the performance of the separation as a function of the \size, rather than the reconstructed energy typically used in stereoscopic analyses.

\subsubsection{Flip determination algorithm}
\label{sec:flip_algo}
In \citet{Murach2015}, the ``flip'' is determined by examining the sign of the \hskew. 
This asymmetry is a good first-order indicator of the orientation of the ellipse with respect to the source position. 
However, for truncated or small images with few pixels and a small \size, the \hskew determination becomes less accurate, resulting in an increasing fraction of incorrectly flipped events.

In this work, we improve the flip determination by employing a boosted decision tree (BDT), a machine-learning algorithm known for its efficiency as a classifier and widely used in \gam-ray astronomy for tasks such as gamma-hadron separation. 
We use the \texttt{XGBRegressor} from the \texttt{XGBoost}\footnote{Version 1.7.4} \texttt{Python} package \citep{xgboost2016} with 100 trees, a maximum depth of 5 and a learning rate of 0.3.

The true flip for a simulated event, $F_{\rm{true}}$, is defined as
\begin{equation}
    F_{\rm{true}} = 
    \begin{cases}
        1,& \ \rm{if } \ X_{\rm{CAM}}^{\rm{COG}} < X_{\rm{CAM}}^{\rm{true}} \\
        -1,& \ \rm{if } \ X_{\rm{CAM}}^{\rm{COG}} > X_{\rm{CAM}}^{\rm{true}}
    \end{cases}
    ,
\end{equation}
with $X_{\rm{CAM}}^{\rm{true}}$ being the true position of the source in the camera reference frame. 
The simulated point-source \gam-ray events were divided into two sets, where we used 80\% for training and 20\% for testing. 
The variables from Sect.~\ref{sec:new-variables} were given as input to the training, and the total fraction of wrongly flipped events after the BDT evaluation served as the score to be minimized.
As BDTs return the relative variable importance for each input feature, we chose the following procedure for the selection of useful variables.

\begin{figure}
    \centering
    \includegraphics[width=.98\linewidth]{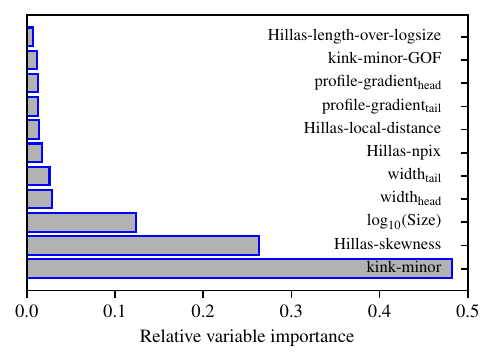}
    \caption{
        Relative variable importance for the flip training.
     }
    \label{fig:flipbdt}
\end{figure}

The selection process began by including all non-time-based variables, followed by an iterative process in which the least important variable was removed and the BDT was re-trained. 
This process continued until the false flip fraction over all energies increased by 2.5\% with respect to the model trained with all variables.
The final set of selected variables and their respective importance are shown in Figure~\ref{fig:flipbdt}.

\subsubsection{Energy and disp regression algorithm}

Energy and ``disp'' regression are currently performed in \hess using two independent shallow neural networks (NNs) based on the \texttt{tensorflow.keras} framework, with two hidden layers with 20 nodes each and a sigmoid activation.
The mean squared error of the predicted variable (energy or disp) for all events was used as the variable loss to be minimized.
The NN were trained for up to 5000 epochs with a batch size of 1000.
90\% of the data were used for training and 10\% for validation, that is, estimation of the variable loss. 
For this work, the structure of the neural networks was kept, changing only the input variables. 
For the standard \hess analyses, the variables used are: \size, \hlength, \hwidth, \hskew, \hkurt, and \hlogdens.
Again, we explored the ideal subset of the variables from Sect.~\ref{sec:new-variables}. 
As the variable importance is not available in NNs, the iterative procedure starts with all variables, conducting a training with $N$ input variables. 
Subsequently, $N$ individual trainings were performed, each omitting one of the $N$ variables, and the final loss of each NN training with the ($N-1$) variables was evaluated.
The best-performing training indicated the least important variable and was used for the next iteration.
This process was repeated until the performance (i.e., variable loss) dropped by 5\% with respect to the model trained with all the variables.
The same subset of variables was chosen for both energy and disp regression: \size, \hlocdist, \hwidth, \hlength, \npix, \hskew, \hkurt, \hlols, \chargeasym, \tgx, and \stda.

\subsubsection{Particle identification algorithm}
Due to the large number of hadronic background events, which exceed the number of \gam-ray events by typically three orders of magnitude, an efficient classification algorithm is necessary.
Its task is to assign a ``gammaness'' score to each event that is maximally different for hadronic and \gam-ray events.
The differences in the gammaness distribution allows the signal to be separated from background events by rejecting events below a certain value of the gammaness score. 
We refer to this value as selection or BDT cut.
Similar to the flip training, we employ a BDT trained on simulated point-source \gam rays and measured background data across all energy and size ranges, with 10\% of the data reserved for validation to prevent over-fitting.
After exploring various settings for the number of trees, tree depth, and learning rate, we identified an optimal configuration with 200 trees, a maximum depth of 6, and a learning rate of 0.3, yielding a stable and robust performance.
We also investigated the impact of segregating the training data by size ranges, but this approach did not result in significant performance improvements.
This indicates that the algorithm has enough information about the event sizes to treat different sizes appropriately.

For the separation training, the events are re-weighted from the simulated spectrum with an index of $-1.8$ to a more typical \gam-ray spectrum with an index of $-2.5$.
We do not expect any bias for sources with different spectral shapes or offsets from the telescope's pointing, as the instrument response functions (IRFs)\footnote{Collection of functions describing the response of the telescope to a true source flux (effective area, energy dispersion, point spread function, and background model)} are binned in energy and offset; however, the performance might decrease slightly.

\begin{figure}
    \centering
    \includegraphics[width=.98\linewidth]{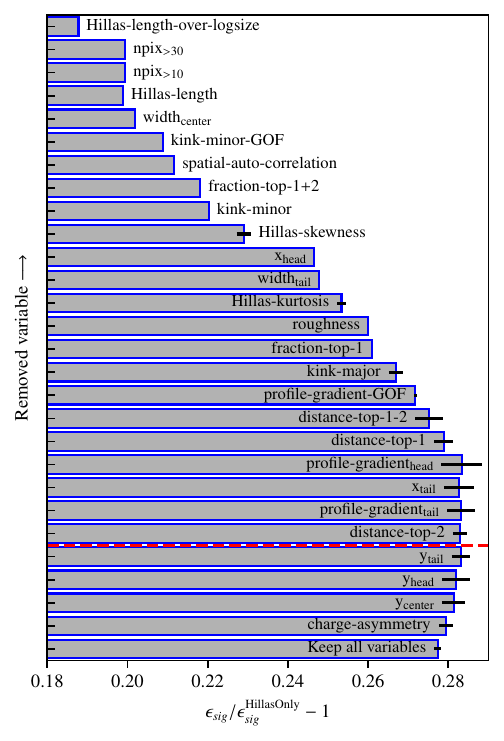}
    \caption{
        Relative change of the signal efficiency $\epsilon_\mathrm{sig}$ during the iterative removal of the least important variable (from bottom to top). The signal efficiency for each training is compared to the signal efficiency of the reference training $\epsilon_\mathrm{sig}^\mathrm{HillasOnly}$ (the current training with the 6 Hillas variables) for a background rejection of 95\%. The black error bars show the difference in the signal efficiency for simulations with nominal and high NSB values, which exceeds the difference present in the reference training. No error bar corresponds to a difference which is smaller than in the reference training. Variables below the dashed red line are not selected for the final training.
     }
    \label{fig:separation-var-selection}
\end{figure}

For the selection of useful variables, we again employ an iterative procedure that starts with all variables (except the time variables) and removes the least important variable after each training iteration.
The performance of each training is evaluated based on the signal efficiency, requiring a background rejection fraction of 95\% over all events.
Additionally, the training is also evaluated on simulations with a high NSB value, and the difference in signal efficiency between the two simulations serves as an indicator of the training's sensitivity to NSB fluctuations.
Figure~\ref{fig:separation-var-selection} summarizes this procedure, where each bar, starting from the bottom, shows the result of the training after removing the respective variable.
The result consists of two parts: the first is the relative improvement in signal efficiency compared to the reference training with the standard six Hillas variables ($\epsilon_\mathrm{sig}^\mathrm{Hillas Only} = 27.8\%$) for the nominal NSB value.
Secondly, the signal efficiency is compared between nominal and high-NSB simulations for all trainings.
The ideal training maximizes improvement in signal efficiency compared to the reference training while not increasing the difference between the two NSB simulation sets.

After removing the first four variables (\chargeasym, \yO, \yI, \yII), the signal efficiency slightly increases, indicating confusion in the BDT due to too many correlated or unimportant variables.
Continuing to remove variables does not affect the performance but increases the NSB sensitivity of the trainings. 
Only after removing \distI do both the performance and the NSB sensitivity decrease, with the difference in signal efficiency between the two NSB simulation sets ending up at values smaller than 1.6\% -- the value for the reference training -- for most of the trainings.
The following variables survived the selection process beyond the iterations shown in Fig.~\ref{fig:separation-var-selection} (in ascending order of their importance): \stda, \kinkxgof, \hlocdist, \npix, \wO, \hlogdens, \hwidth, and \pg.

Mainly because of the sensitivity to different NSB rates, we decided to keep all variables except \chargeasym, \yO, \yI, \yII, although other selections could be justified as well.
For the final training we remain with 31 input variables, with their relative variable importance given in Tab.~\ref{tab:var_summary}.

\subsection{Size-dependent selection cut}
\label{sec:size-dependent-sel-cut}
\begin{figure}
    \centering
    \includegraphics[width=.98\linewidth]{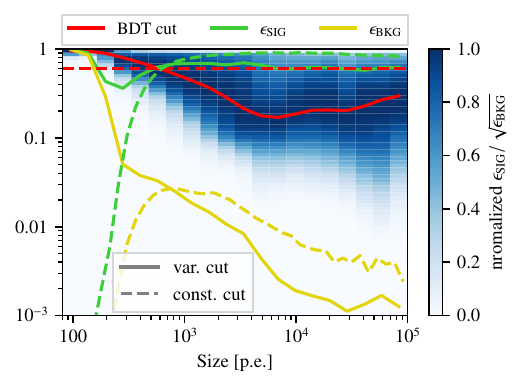}
    \caption{
        Scan of q-factors for different event sizes (x-axis) and BDT cuts (y-axis) using the 6 Hillas variables. For each size bin the color map shows the q-factors (normalized to 1) for different BDT cuts. The resulting optimal BDT cut (red) is shown together with the signal (green) and background (yellow) background efficiencies. The solid lines show the case for the size-dependent BDT cut while the dashed lines show the constant BDT cut.
     }
    \label{fig:bdt-scan}
\end{figure}

To find the optimal BDT cut resulting in the best sensitivity, one can maximize the q-factor defined as $q = \epsilon_\mathrm{sig} / \sqrt{\epsilon_\mathrm{bkg}}$.
This approach is effective because the observed significance $\sigma$ for an on-off (i.e., aperture photometry) analysis is proportional to 
\begin{equation}
    \sigma \propto \frac{N_\mathrm{sig}}{\sqrt{N_\mathrm{sig} + N_\mathrm{bkg}}} = \frac{n_\mathrm{sig}\cdot\epsilon_\mathrm{sig}}{\sqrt{n_\mathrm{sig}\cdot\epsilon_\mathrm{sig} + n_\mathrm{bkg}\cdot\epsilon_\mathrm{bkg}}} \approx \frac{n_\mathrm{sig}}{\sqrt{n_\mathrm{bkg}}} \frac{\epsilon_\mathrm{sig}}{\sqrt{\epsilon_\mathrm{bkg}}}\,,
\end{equation}
where $N$ refers to counts after cuts and $n$ refers to the number of events before cuts.
The final approximation assumes $N_\mathrm{sig} << N_\mathrm{bkg}$ which is a good approximation in the low signal-to-noise case where optimization is most critical.
The ratio of $n_\mathrm{sig}/\sqrt{n_\mathrm{bkg}}$ is fully determined by the source and background spectra together with the effective area, leaving the q-factor as the remaining quantity to optimize.
Several tests carried out in the context of this study showed that the performance of the BDT within a certain size range is stable with respect to adding or removing events of different sizes\footnote{This holds as long as the algorithm has access to (implicit) information about the event size.}. 
However, the distribution of the BDT score changes, which can make the performance appear worse or better for the size range of interest if the BDT cut is optimized using events of all sizes. 
Therefore, we optimize the q-factor in different size bins by scanning through different BDT cuts (cf. Fig.~\ref{fig:bdt-scan}).
The optimal BDT cuts are then smoothed with a Gaussian kernel with width equal to one bin, which results in the red line in Fig.~\ref{fig:bdt-scan}.
Events can now be selected based on the BDT cut interpolated at their respective size and their BDT score.

This approach allows for the retention of more low-size events, where due to poor separation, the best statistical significance is achieved by keeping nearly all events. 
While the sensitivity gain due to low-size events is limited by the high background contamination, it does lower the energy threshold and reduces the relative systematic uncertainty, which is highest for strict cuts in the distribution.

\section{Performance evaluation}
\label{sec:performance}
The performance evaluation is conducted for simulations at a zenith angle of $20^\circ$ and an angular separation (offset) of $0.5^\circ$ from the pointing direction of the telescopes.
In the following, we compare the current standard analysis using the six Hillas variables and a constant BDT cut (labeled ``Hillas variables (const. cut)'') to an intermediate analysis with the Hillas variables but a size-dependent BDT cut (labeled ``Hillas variables (var. cut)'') and the new analysis with new variables and the size-dependent cut (labeled ``New variables (var. cut)'').
Details on how the size-dependent BDT cut is computed can be found in Sect.~\ref{sec:size-dependent-sel-cut}.

The performance presented in \citet{Murach2015} is based on different event (pre)selection cuts and the camera previously installed on CT5.
In contrast to this study, they use a $\theta^2$-cut on the angular mis-reconstruction, considerably lowering the effective area but improving the energy resolution.
For these reasons we do not directly compare to the results from \citet{Murach2015}, but instead to the performance of current FlashCam analyses without $\theta^2$-cut. 

\subsection{Event reconstruction}
\label{sec:event-reco}
In this section, we evaluate the performance of the event reconstruction using the machine learning models described in Sect.~\ref{sec:architecture}.
First, the ``flip'' is determined, based on which the sign of variables like the \hskew is changed, and sector variables like \xO and \xII are interchanged according to Tab.~\ref{tab:var_summary}.
The flipped variables then serve as input for the energy, direction, and particle type reconstruction.

\subsubsection{False flip fractions}
\label{sec:fff}
\begin{figure}[h]
    \centering
    \includegraphics[width=.98\linewidth]{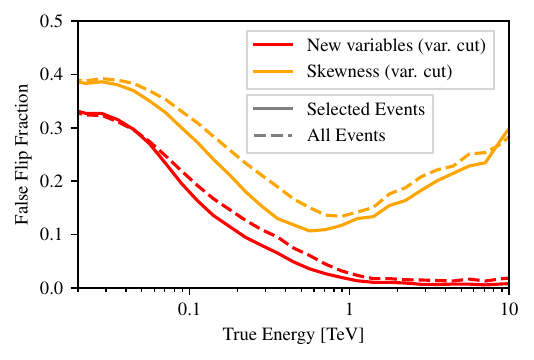}
    \caption{
        Fraction of wrongly flipped events as function of true energy.
        The dashed lines show the performance evaluated on all events after preselection, while the solid lines only include selected events passing the size-dependent gamma-hadron separation cut.
     }
    \label{fig:fff}
\end{figure}

The improvements in the flip determination are shown in Fig.~\ref{fig:fff}, where the performance of the BDT model is compared with that of the flip determination based solely on the \hskew.
At high energies, where an increasing fraction of the events is truncated at the camera edges, image skewness alone becomes an unreliable indicator.
In contrast, the BDT model consistently demonstrates the expected behavior, with a decreasing false flip fraction as energy increases.
For low-energy events with small images, determining the correct flip remains challenging.
This is especially the case for highly symmetric events with their CoG close to the true source position.
However, for these events, even an incorrect flip has only a minor effect on the reconstructed position and the associated parameters.
As a result, the false flip fraction merely counts all events where the flip was incorrectly determined, while the angular resolution more accurately reflects the impact of the flip determination on the reconstruction.

\subsubsection{Angular resolution}
The angular resolution is defined as the 68\% containment radius of the reconstructed position relative to the true one.
It is important when disentangling sources close to each other and also affects the sensitivity of the instrument to isolated point-like sources, as the background increases with a larger integration area of the signal.

\begin{figure}[h]
    \centering
    \includegraphics[width=.98\linewidth]{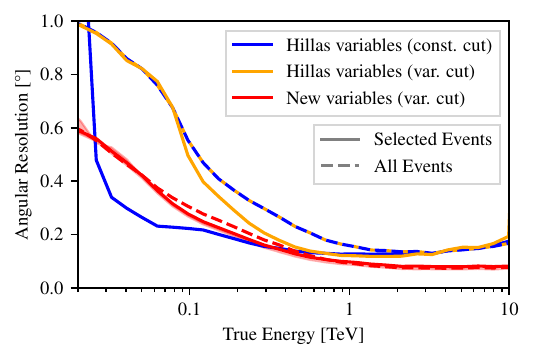}
    \caption{
        Angular resolution as a function of true energy.
        The blue curves show the performance for the current analysis, the orange ones for the intermediate analysis with the old training but the size-dependent BDT cut, and the red lines represent the new analysis. The red-shaded band shows the difference to high-NSB simulations.
     }
    \label{fig:angres}
\end{figure}

Based on the dashed lines in Fig.~\ref{fig:angres}, the improvement achieved with the new variables is evident, lowering the angular resolution at 100\,GeV by 57\%.
The dashed lines are always shown to represent the performance of the reconstruction independent of the BDT cut, while the solid lines represent the actual performance of the analysis including only events passing the gamma-hadron separation.
Since the constant BDT cut removes most low-size events, the good angular resolution of this analysis at low energies is a natural but mostly irrelevant consequence, due to the low event statistics in this range.
The enhancement in angular resolution is partly attributable to the improved ``disp'' estimation and partly to the improved flip training, as incorrectly flipped events are often reconstructed far from the true position, particularly when the impact distance is large.

As the direction determination of the new analysis also makes use of very NSB dependent variables such as the \tgx, the angular resolution is also evaluated on the high-NSB simulations.
Here, we use simulations of \gam-ray events with an increased NSB value (scaled by 1.65), which is higher than expected for the vast majority of observations (cf. Sect.~\ref{sec:AD-test}).
The difference in angular resolution is very small and is shown in the barely visible red-shaded region in Fig.~\ref{fig:angres}.

\subsubsection{Energy resolution}
\begin{figure}[h]
    \centering
    \includegraphics[width=.98\linewidth]{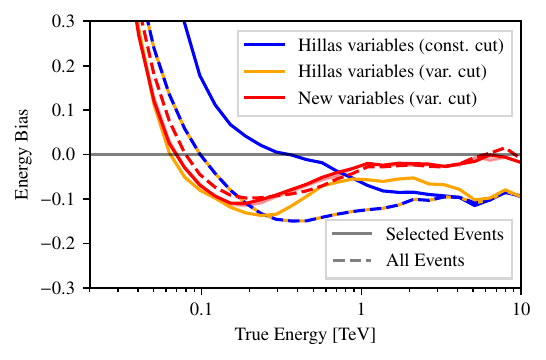}
    \caption{
        Energy bias as a function of true energy.
     }
    \label{fig:ebias}
\end{figure}

The energy bias is defined as the mean of the energy migration $\mu = (E_\mathrm{reco}-E_\mathrm{true})/E_\mathrm{true}$ and the energy resolution is the standard deviation of $\mu$.
For ideal reconstruction, the energy bias would be zero; however, if events with different true energies appear very similar to the algorithm, it assigns an average energy to minimize the loss. 
This leads to a positive bias for low-energy events and a negative bias for higher-energy events.
The dip-like shape of the curves for a variable BDT cut in Fig.~\ref{fig:ebias} can be understood when considering that events with low size are composed of events with low energies and small impact distance, as well as events with large energies and large impact distance.
Compared to the other curves, the energy bias with the constant BDT cut appears shifted to the right, as only very few events with low sizes pass the BDT cut.
The energy at which the bias exceeds a value of +10\% is usually considered a safe choice for the energy threshold.
Hence, we observe that the energy threshold can be lowered considerably using the size-dependent BDT cut.
Comparing the dashed lines, for which no BDT cut is applied, one can see that the new variables help in moving the energy bias closer to zero.

\begin{figure}[h]
    \centering
    \includegraphics[width=.98\linewidth]{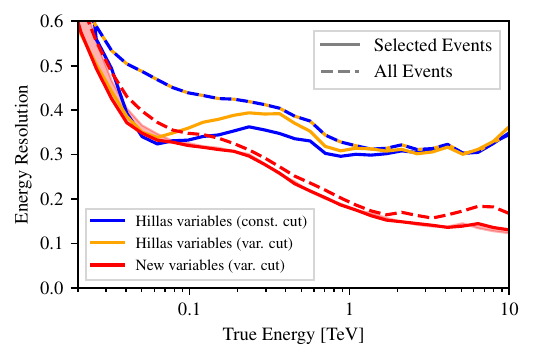}
    \caption{
        Energy resolution as a function of true energy. The red-shaded region shows the difference in energy resolution compared to the high-NSB simulations.
     }
    \label{fig:eres}
\end{figure}

The energy resolution, shown in Fig.~\ref{fig:eres}, also shows significant improvement when evaluated on all events.
Considering the BDT cuts, the improvement is most notable at higher energies $\gtrsim$200\,GeV.

The difference to high-NSB simulations is again illustrated based on the red-shaded regions. 
Although there is no significant difference in the energy bias, the energy resolution becomes clearly worse for low energies under high-NSB conditions. 
This is not unexpected, as during calibration, the pixel amplitudes are corrected on the basis of their average amplitude in the absence of a triggered event, leading to a minor effect of the NSB rates on the energy bias.
However, the NSB fluctuations in the pixel amplitudes increase with the square root of the subtracted amplitude, causing also larger uncertainty in the energy reconstruction.

\subsection{Background rejection}
\label{sec:bkg-rejection}

\begin{figure}
    \centering
    \includegraphics[width=.98\linewidth]{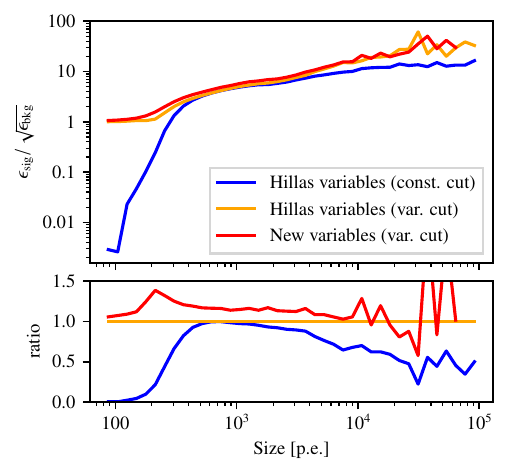}
    \caption{
        The absolute q-factors (upper panel) of the separation trainings as a function of event size, and their ratios (lower panel).
        The blue curves show the performance for the current analysis, the orange ones for the intermediate analysis with the old training but the size-dependent BDT cut, and the red lines represent the new analysis.
     }
    \label{fig:q-factors}
\end{figure}

The difference in separation performance between the three configurations, measured in terms of the q-factor (see Sect.~\ref{sec:size-dependent-sel-cut}), is shown in Fig.~\ref{fig:q-factors}.
Here, the constant BDT cut is optimized for medium sizes around 600\,p.e., resulting in similar q-factors for the two configurations that use the standard Hillas variables.
At lower or higher sizes the configurations employing the size-dependent cut achieve significantly higher q-factors.
The benefit of the new variables is most pronounced for sizes around 200\,p.e. (not considering the statistical fluctuations at larger sizes).

\begin{figure}
    \centering
    \includegraphics[width=.98\linewidth]{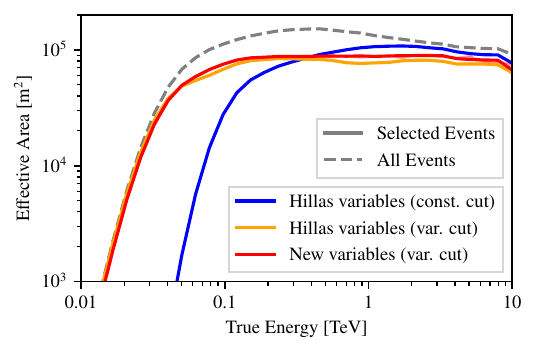}
    \caption{
        Effective areas for the three configurations as function of true energy.
     }
    \label{fig:aeff}
\end{figure}

Finally, the effective areas shown in Fig.~\ref{fig:aeff} also reflect the lower energy threshold for the configurations with a variable BDT cut, which is also often chosen as the energy at which the effective area drops below 10\% of its maximum.
This threshold is at $30$\,GeV for the configurations with a variable BDT cut and at $64$\,GeV for the configuration with the constant BDT cut.

\begin{figure}
    \centering
    \includegraphics[width=.98\linewidth]{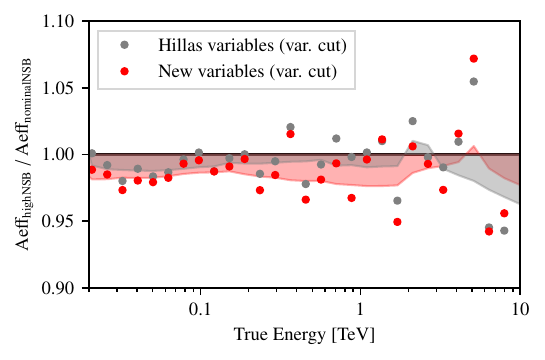}
    \caption{
        Ratio of the effective areas computed from simulations with nominal and high NSB values. The points show the ratio for each energy bin while the shaded bands represent an estimate of the systematic uncertainty. For details see main text.
     }
    \label{fig:aeff-sys-nsb}
\end{figure}

In a real data analysis with the HAP framework, the corresponding IRFs are computed based on simulations at different zenith, azimuth, and offset angles, with atmospheric transparency also accounted for.
One variable not considered is the rate of the night sky background light.
In Fig.~\ref{fig:aeff-sys-nsb}, the resulting effective area for the high-NSB simulations is compared to that from the default simulations.
For observations taken under such high NSB rates, this difference would translate directly into a systematic error on the recovered flux.
Since the number of events simulated for the high NSB value is considerably lower than for the default simulations, the ratio of the effective area is also affected by statistical fluctuations, especially toward the higher energies.

The shaded bands in Fig.~\ref{fig:aeff-sys-nsb} are constructed based on a systematic deviation of neighboring points from a value of 1 and therefore still include statistical fluctuations.
Especially toward higher energies, the statistical fluctuations dominate and increase the difference between the two simulation sets.
The width of the shaded bands can thus be interpreted as the maximum expected systematic uncertainty caused by different NSB values.
This systematic effect resulting from high NSB rates is expected to be opposite for observations taken under lower NSB rates, although less strong.

Quantitatively, the systematic uncertainty due to NSB differences is $\sim$2.5\%, which is small compared to other systematic uncertainties, which are generally estimated to add up to $\sim$20\% \citep[see supplementary material of][]{rsoph-2022}.
However, it is still informative to compare this systematic uncertainty between the new and the old variables.
This comparison is made for the two configurations with the size-dependent BDT cut, as these have similar effective areas, and demonstrates that the systematic uncertainty is slightly increased by approximately one percentage point.
This suggests that the increased performance in gamma-hadron separation comes at the expense of a slight increase in susceptibility to data-simulation mismatches, as for example introduced by different NSB values.

\subsection{Sensitivity}
\label{sec:sensitivity}

Using all the IRFs, we compute sensitivities for source detection in different energy bins.
In each energy bin, the source flux ($\propto E^{-2.5}$) is scaled such that it leads to a 5$\sigma$ detection for 50\,h of observation time.
Since the absolute values of this differential sensitivity depend on the size of the energy bins, they primarily serve as a means of comparison between different configurations.

Scanning through various preselection cuts for image size and pixel number showed that small events, with five or six pixels, generally decrease the sensitivity.
These events exhibit low resolution during reconstruction and perform especially poorly in gamma-hadron separation.
Including such events in the analysis significantly increases the background without a comparable enhancement in the signal.
For this reason, the following analysis is based on slightly stricter preselection criteria, where events must contain seven or more pixels. 
While applying even more stringent cuts could further enhance performance at higher energies, this comes at the cost of diminished performance at lower energies, which is the primary focus of this study. 

\begin{figure}
    \centering
    \includegraphics[width=.98\linewidth]{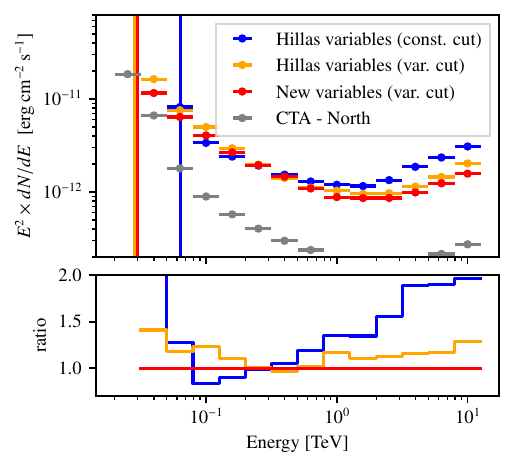}
    \caption{
        Comparison of the differential sensitivities (upper panel) and their ratios (lower panel). The vertical lines show the energy threshold at which the effective area drops below 10\% of its maximum. The sensitivities are computed for 50\,h of observation time and 5$\sigma$ detection per energy bin. The points for CTA - North rely on prod5 v0.1 IRFs\protect\footnotemark.
     }
    \label{fig:sens-comp}
\end{figure}

Figure~\ref{fig:sens-comp} highlights the improvement in sensitivity, which roughly follows the improvement in q-factor already discussed in Fig.~\ref{fig:q-factors}.
The constant BDT cut is optimized for medium energies around 100\,GeV, where due to the strict cut on low-size events, the sensitivity is slightly better compared to the new configurations.
Improvements from the variable cut are apparent at higher and lower energies.
The new variables mainly increase the performance in the first few bins, where they mostly contribute to the increased background rejection and improved event reconstruction.
At the threshold, an improvement of 41\% could be achieved by adding the new variables.

\footnotetext{taken from: \url{https://www.ctao.org/for-scientists/performance/}}

\subsection{Application to test dataset}
\label{sec:data}
In this section, we apply the new reconstruction to 65 observations ($\sim\,$30\,h) of the Crab Nebula, taken under zenith angles between 44$^\circ$ and 55$^\circ$.
The dataset used is identical to that in \citet{crab-hess-2024} for the mono data, where the standard reconstruction technique was applied.
For the current study, the background is estimated using reflected regions around the source position \citep{Berge2007}.
A low-energy threshold of 100\,GeV is chosen, which lies above the 10\% effective area threshold and below the 10\% energy bias threshold for this zenith range. 
At this energy threshold, the new analysis exhibits an energy bias of 41\%, whereas the previous analysis demonstrated a bias of 128\%. 
At the threshold energy of the previous analysis, 200\,GeV, the new analysis shows a significantly reduced bias of 0.6\%, compared to 27\% for the old analysis.

\begin{figure}
    \centering
    \includegraphics[width=.98\linewidth]{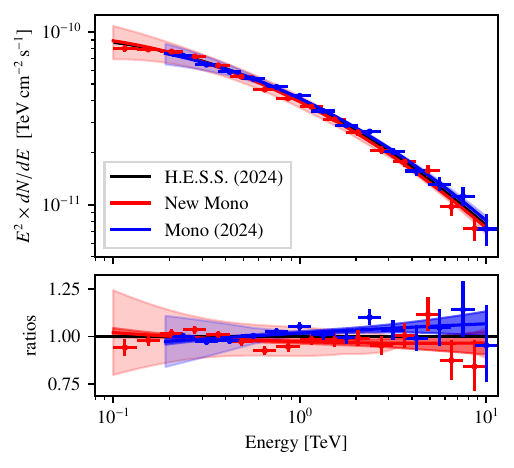}
    \caption{
        Spectral energy distribution of the Crab Nebula as measured by CT5. The blue data points correspond to the \hess mono data from \citet{crab-hess-2024}, reflecting the performance of the standard reconstruction method. The black line represents the Crab spectrum obtained from the mono + stereo analysis (extrapolated to 100\,GeV) also from \citet{crab-hess-2024}. The red data points indicate the results from the new reconstruction method. In the upper panel, the shaded error bands represent the combined statistical and systematic uncertainties. The lower panel illustrates the ratios relative to the black spectrum, with the darker shaded regions showing the statistical uncertainty in addition to the total uncertainty. 
     }
    \label{fig:crab-spec}
\end{figure}

Figure~\ref{fig:crab-spec} provides a comparative overview of the spectra derived from the two analyses. 
The systematic uncertainty is estimated by grouping the observations into three different zenith and offset angle bins, respectively, and estimating the deviations beyond the statistical uncertainty.
This kind of systematic can be regarded as lower limit of systematic uncertainties, particularly reliable at lower energies where statistical uncertainties are minimal.
Overall, we find good agreement between the old and new analyses, with the latter extending to lower energies.
As expected, systematic uncertainties increase at lower energies, primarily due to the inclusion of events near the instrument's trigger threshold.

\section{Conclusion}
In this study, we introduced approximately 30 new image parameters and assessed their impact on the reconstruction and gamma-hadron separation of monoscopic IACT images.
One of the primary challenges for monoscopic events is determining the correct image flip, that is, whether the arrival direction is to the left or right of the image's center of gravity. 
By employing a boosted decision tree for this task, we improved the accuracy of the flip determination by approximately eight percent points up to 1 TeV, with even larger improvements at higher energies.
The fraction of wrongly flipped events could be reduced from 38\% to 32\% at threshold and from 29\% to 17\% at 100\,GeV.
The most important parameter for the BDT was the new \kinky variable, which describes the change in image width along the major shower axis.

Unlike the traditional Hillas parameters, many of the new parameters lack symmetry with respect to the image center. 
Consequently, if the image is mirrored along the y-axis, the sign of certain variables may change, or variables associated with the head and tail regions may be swapped. 
Correctly determining the flip is therefore crucial not only for angular reconstruction but also for the accurate interpretation of these new variables.
Throughout this study we put great emphasis on removing any asymmetries between the distributions of \gam-ray and background events by consistently mirroring half of the events.
This approach ensures uniform performance regardless of the event's position within the camera frame, while fully leveraging the information provided by the image parameters. 
As a result, both the energy and angular resolutions showed considerable improvement.
Additionally, the gamma-hadron separation saw a notable enhancement, with an average 20\% increase in the q-factor for low to medium-size events. 

When combined with the new size-dependent BDT cut, these reconstruction improvements allow for the inclusion of more low-energy events in the analysis, significantly lowering the energy threshold. 
For observations at a zenith angle of 20$^\circ$ and an offset of 0.5$^\circ$, the 10\% energy bias threshold was reduced from 120 GeV to 50 GeV, and the 10\% effective area threshold from 64 GeV to 30 GeV.
Although background rejection at low energies remains a challenge, the sensitivity curves presented in this work are promising.
We also carefully evaluated the systematic uncertainties introduced by varying NSB rates and found only a minor increase in the systematic error on the effective area based on the new image parameters.

These three improvements to the analysis of monoscopic events were successfully applied to \hess data, demonstrating their potential for integration into the analysis of other IACT arrays.
Like the \hess array, the upcoming Cherenkov Telescope Array Observatory (CTAO) will feature telescopes of various sizes, with only the largest telescopes detecting the faint light signals from low-energy \gam rays.
Consequently, the relevance of monoscopic event analysis will persist, even with the superior performance of stereoscopic events.
In fact, mono- and stereoscopic events may be classified into distinct event types, allowing for a combined analysis of all events without sacrificing sensitivity.

Future studies could explore the analysis of IACT data split into different event-types and the potential of time-cleaning for the shower images observed with FlashCam.
The latter might reduce the susceptibility of the time-related parameters to NSB fluctuations, enabling their inclusion in the background rejection algorithms, and further minimizing systematic uncertainties.
Template-based reconstruction methods for monoscopic events offer additional potential for enhancing event reconstruction; however, their performance is highly dependent on the quality of the initial seeds.
In this context, the improvements presented in this work are expected to provide substantial value.

\section{Data availability}
Appendix \ref{appx:new-var} is available on Zenodo (\url{https://doi.org/10.5281/zenodo.14640478}).

\begin{acknowledgements}
We thank the \hess Collaboration for providing the simulated data, common analysis tools, and valuable comments to this work. 
We also thank the \hess Collaboration for allowing us to use the data on the Crab nebula in this publication.
This research made use of the \textsc{Astropy} (\url{https://www.astropy.org}; \citealt{Astropy2013,Astropy2018,Astropy2022}), \textsc{matplotlib} (\url{https://matplotlib.org}; \citealt{Hunter2007}), \textsc{iminuit} (\url{https://iminuit.readthedocs.io}; \citealt{Dembinski2020}) and \textsc{gammapy} (\url{https://gammapy.org/}; \citealt{gammapy:2023, gammapy:zenodo-1.2}) software packages.
\end{acknowledgements}

\bibliographystyle{aa}
\bibliography{references}

\begin{appendix}
\newpage

\section{Variable distribution plots}
\label{appx:new-var}
For each of the variables we show the distribution of $\sim$\,\num{7e6} background events taken from observations where known \gam-ray sources are excluded. 
We compare this distribution to the distribution of $\sim$\,\num{8e4} simulated protons.
Additionally, we show the distributions of $\sim$\,\num{4e5} diffuse \gam rays simulated with the realistic NSB value and a NSB value which is $\sim$\,1.65 times higher.
All simulations are done for \SI{20}{deg} zenith angle and the simulated spectrum $\sim E^{-1.8}$ is re-weighted to $E^{-2.5}$.
The distributions are normalized to 1 such that the difference in event numbers is not visible which means the y-axis of the plots shows the normalized number of entries in that bin.
The correct image orientation is determined based on the \textit{Flip}-BDT described in Sect.~2.2.1 of the main paper, leading to asymmetries between the head and tail sectors.

\subsection{Hillas parameter-based variables}

\begin{center}
    \includegraphics[width=.98\linewidth]{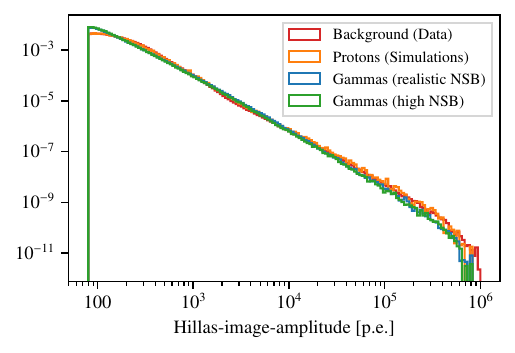}
    \captionof{figure}{
        Distribution of the \size. Protons and gammas are simulated, while background events are measured data.
     }
    \label{fig:dist-size}
\end{center}
\begin{center}
    \includegraphics[width=.98\linewidth]{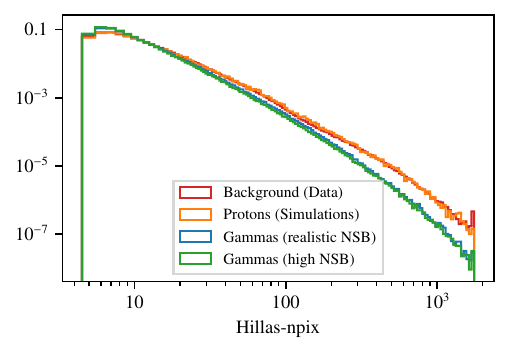}
    \captionof{figure}{
        Distribution of the \npix.
     }
    \label{fig:dist-npix}
\end{center}
\begin{center}
    \includegraphics[width=.98\linewidth]{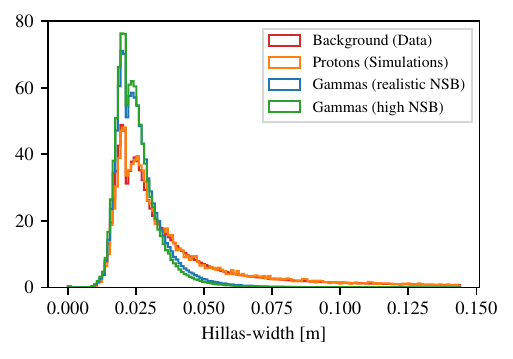}
    \captionof{figure}{
        Distribution of the \hwidth.
     }
    \label{fig:dist-width}
\end{center}
\begin{center}
    \includegraphics[width=.98\linewidth]{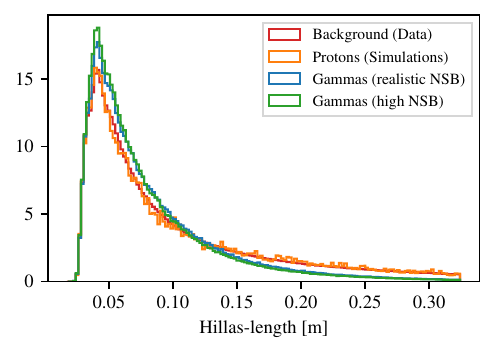}
    \captionof{figure}{
        Distribution of the \hlength.
     }
    \label{fig:dist-length}
\end{center}
\begin{center}
    \includegraphics[width=.98\linewidth]{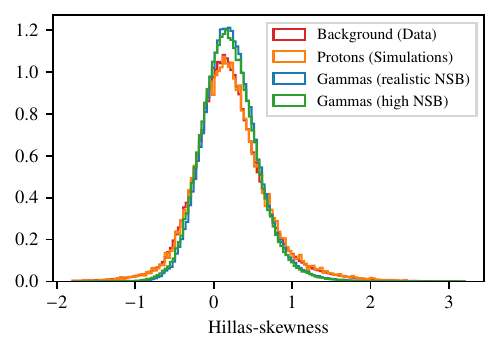}
    \captionof{figure}{
        Distribution of the \hskew.
     }
    \label{fig:dist-skew}
\end{center}
\begin{center}
    \includegraphics[width=.98\linewidth]{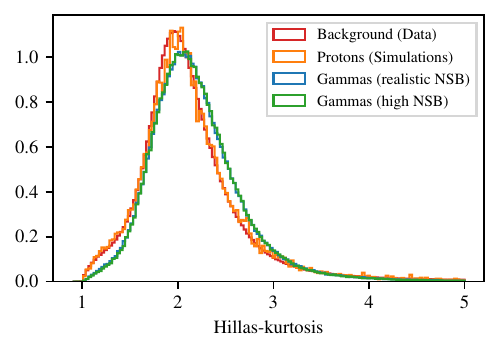}
    \captionof{figure}{
        Distribution of the \hkurt.
     }
    \label{fig:dist-kurt}
\end{center}
\begin{center}
    \includegraphics[width=.98\linewidth]{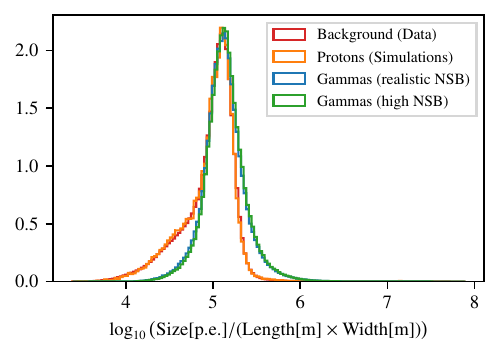}
    \captionof{figure}{
        Distribution of the \hlogdens.
     }
    \label{fig:dist-logdens}
\end{center}
\begin{center}
    \includegraphics[width=.98\linewidth]{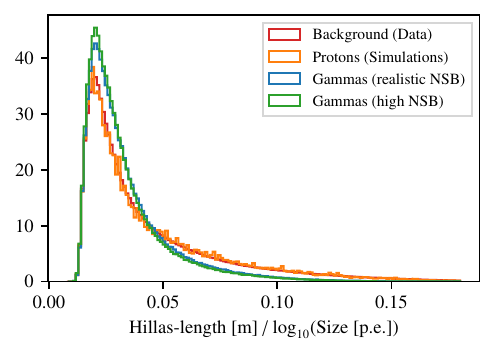}
    \captionof{figure}{
        Distribution of the \hlols.
     }
    \label{fig:dist-lols}
\end{center}

\subsection{Sector-based variables}
\begin{center}
    \includegraphics[width=.98\linewidth]{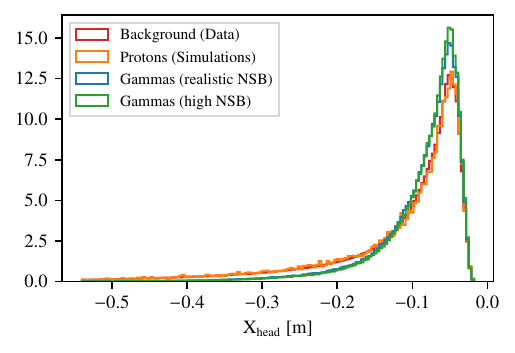}
    \captionof{figure}{
        Distribution of the \xII.
     }
    \label{fig:dist-xO}
\end{center}
\begin{center}
    
    \includegraphics[width=.98\linewidth]{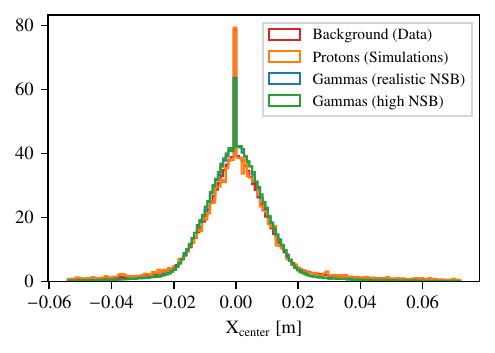}
    \captionof{figure}{
        Distribution of the \xI.
     }
    \label{fig:dist-xI}
\end{center}
\begin{center}
    
    \includegraphics[width=.98\linewidth]{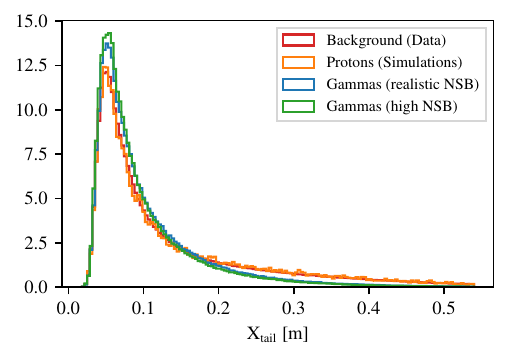}
    \captionof{figure}{
        Distribution of the \xO.
     }
    \label{fig:dist-xII}
\end{center}
\begin{center}
    
    \includegraphics[width=.98\linewidth]{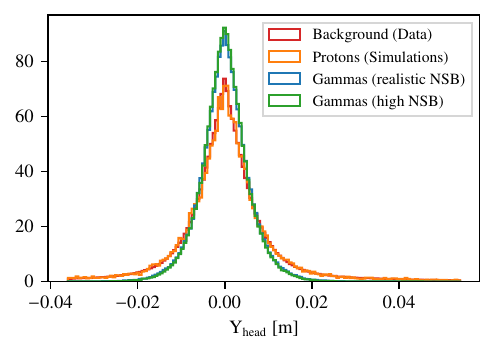}
    \captionof{figure}{
        Distribution of the \yO.
     }
    \label{fig:dist-yO}
\end{center}
\begin{center}
    
    \includegraphics[width=.98\linewidth]{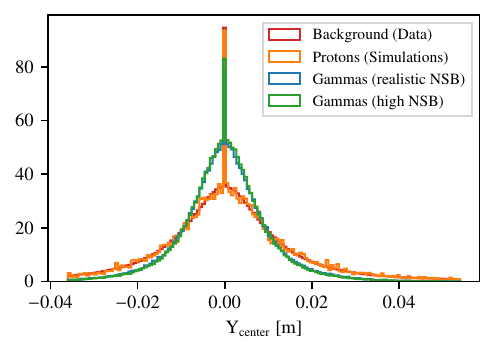}
    \captionof{figure}{
        Distribution of the \yI.
     }
    \label{fig:dist-yI}
\end{center}
\begin{center}
    
    \includegraphics[width=.98\linewidth]{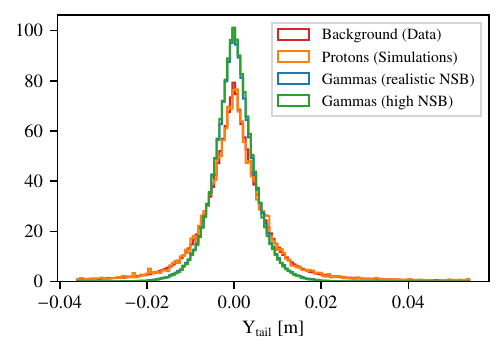}
    \captionof{figure}{
        Distribution of the \yII.
     }
    \label{fig:dist-yII}
\end{center}
\begin{center}
    
    \includegraphics[width=.98\linewidth]{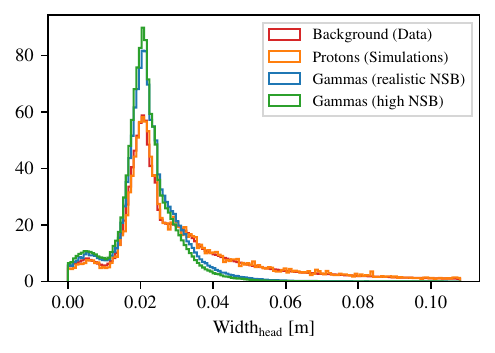}
    \captionof{figure}{
        Distribution of the \wO.
     }
    \label{fig:dist-wO}
\end{center}
\begin{center}
    
    \includegraphics[width=.98\linewidth]{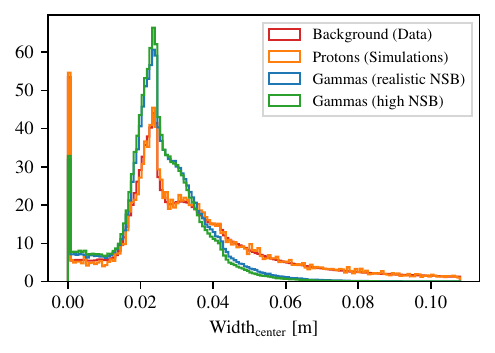}
    \captionof{figure}{
        Distribution of the \wI.
     }
    \label{fig:dist-wI}
\end{center}
\begin{center}
    
    \includegraphics[width=.98\linewidth]{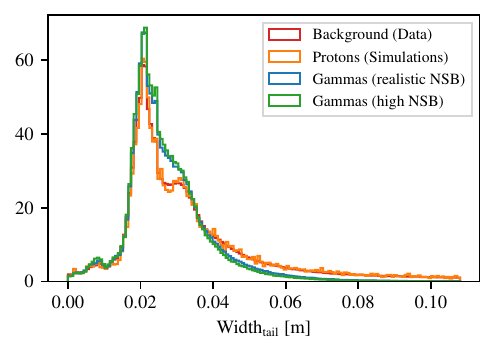}
    \captionof{figure}{
        Distribution of the \wII.
     }
    \label{fig:dist-wII}
\end{center}
\begin{center}
    
    \includegraphics[width=.98\linewidth]{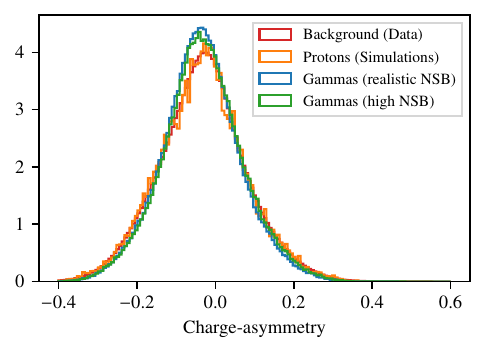}
    \captionof{figure}{
        Distribution of the \chargeasym.
     }
    \label{fig:dist-chargeasym}
\end{center}

\subsection{Continuous variables}
\begin{center}
    
    \includegraphics[width=.98\linewidth]{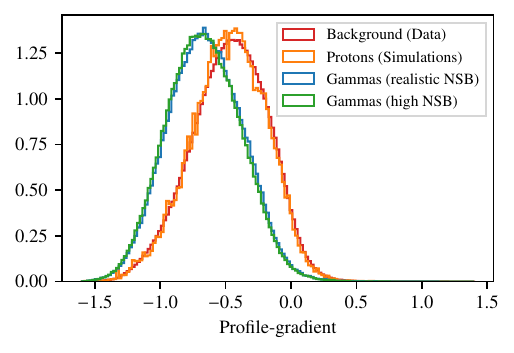}
    \captionof{figure}{
        Distribution of the \pg.
     }
    \label{fig:dist-pg}
\end{center}
\begin{center}
    
    \includegraphics[width=.98\linewidth]{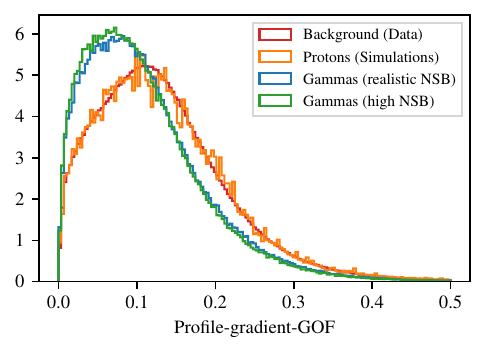}
    \captionof{figure}{
        Distribution of the \pggof.
     }
    \label{fig:dist-pggof}
\end{center}
\begin{center}
    
    \includegraphics[width=.98\linewidth]{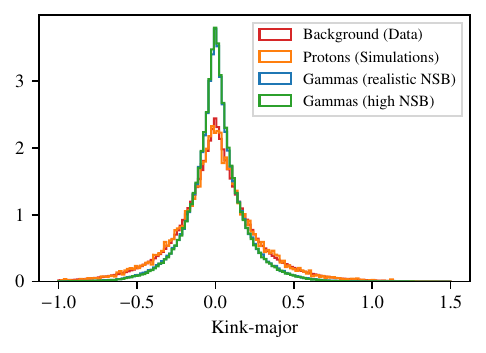}
    \captionof{figure}{
        Distribution of the \kinkx.
     }
    \label{fig:dist-kinkx}
\end{center}
\begin{center}
    
    \includegraphics[width=.98\linewidth]{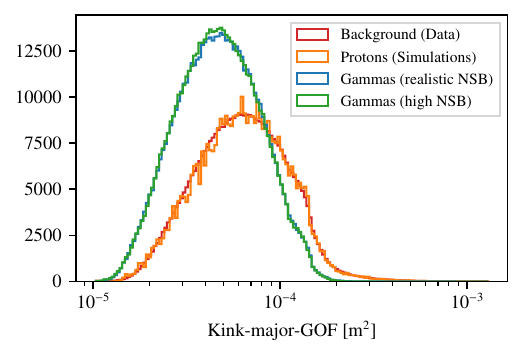}
    \captionof{figure}{
        Distribution of the \kinkxgof.
     }
    \label{fig:dist-kinkxgof}
\end{center}
\begin{center}
    
    \includegraphics[width=.98\linewidth]{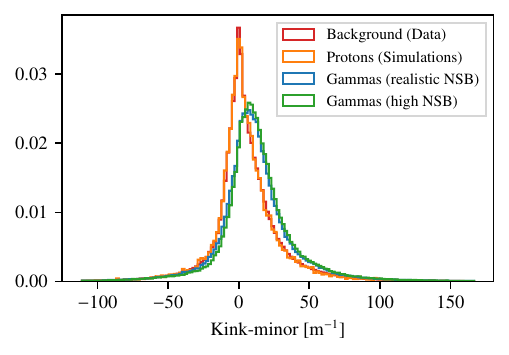}
    \captionof{figure}{
        Distribution of the \kinky.
     }
    \label{fig:dist-kinky}
\end{center}
\begin{center}
    
    \includegraphics[width=.98\linewidth]{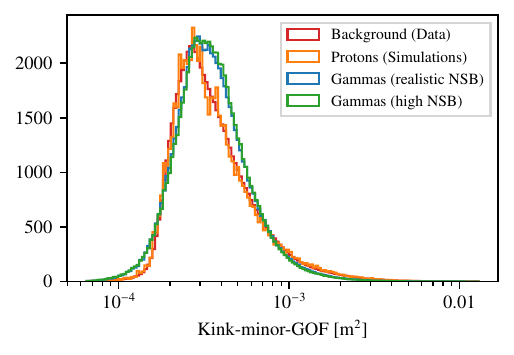}
    \captionof{figure}{
        Distribution of the \kinkygof.
     }
    \label{fig:dist-kinkygof}
\end{center}
\begin{center}
    
    \includegraphics[width=.98\linewidth]{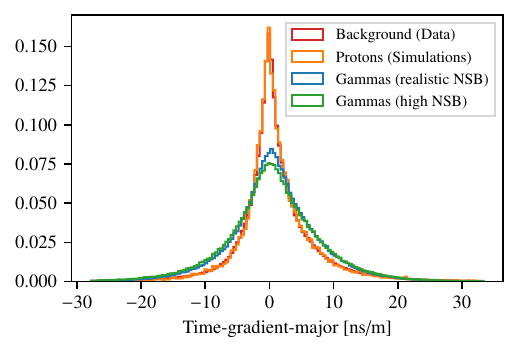}
    \captionof{figure}{
        Distribution of the \tgx.
     }
    \label{fig:dist-tgx}
\end{center}
\begin{center}
    
    \includegraphics[width=.98\linewidth]{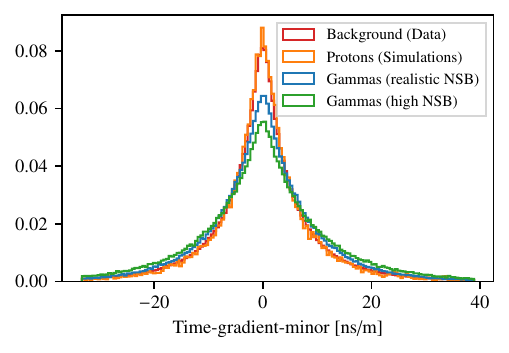}
    \captionof{figure}{
        Distribution of the \tgy.
     }
    \label{fig:dist-tgy}
\end{center}
\begin{center}
    
    \includegraphics[width=.98\linewidth]{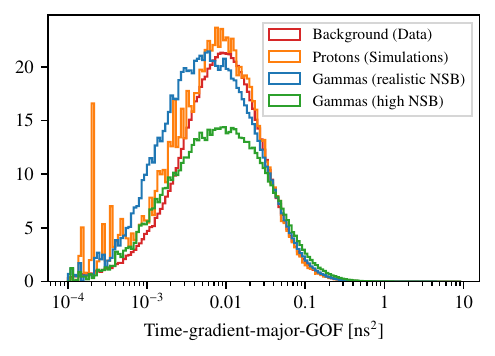}
    \captionof{figure}{
        Distribution of the \tgxgof.
     }
    \label{fig:dist-tgxgof}
\end{center}
\begin{center}
    
    \includegraphics[width=.98\linewidth]{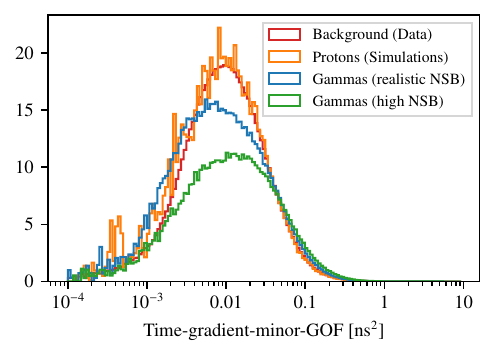}
    \captionof{figure}{
        Distribution of the \tgygof.
     }
    \label{fig:dist-tgygof}
\end{center}
\begin{center}
    
    \includegraphics[width=.98\linewidth]{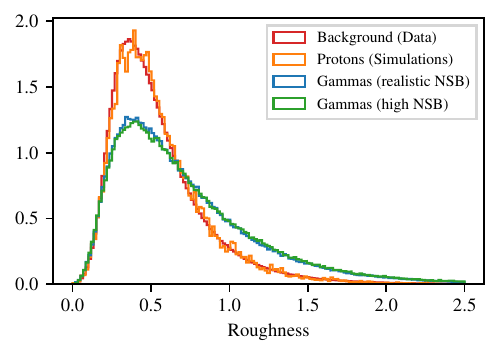}
    \captionof{figure}{
        Distribution of the \roughI.
     }
    \label{fig:dist-rough1}
\end{center}
\begin{center}
    
    \includegraphics[width=.98\linewidth]{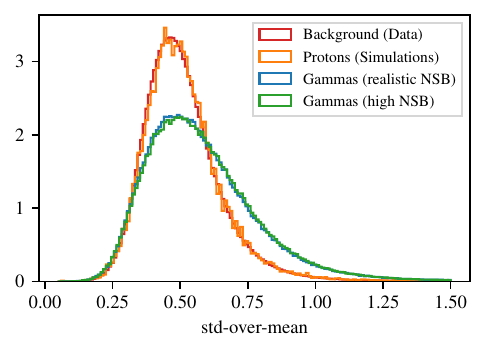}
    \captionof{figure}{
        Distribution of the \stda.
     }
    \label{fig:dist-stda}
\end{center}
\begin{center}
    
    \includegraphics[width=.98\linewidth]{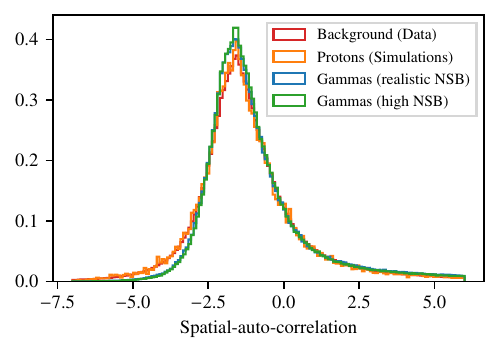}
    \captionof{figure}{
        Distribution of the \roughII.
     }
    \label{fig:dist-rough2}
\end{center}

\subsection{Pixel Variables}
\begin{center}
    
    \includegraphics[width=.98\linewidth]{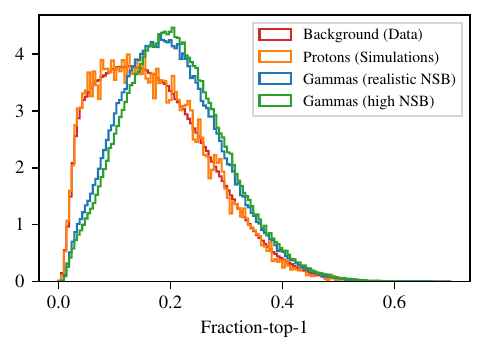}
    \captionof{figure}{
        Distribution of the \ftI.
     }
    \label{fig:dist-ft1}
\end{center}
\begin{center}
    
    \includegraphics[width=.98\linewidth]{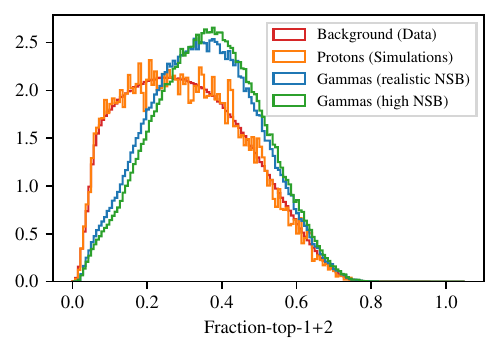}
    \captionof{figure}{
        Distribution of the \ftII.
     }
    \label{fig:dist-ft2}
\end{center}
\begin{center}
    
    \includegraphics[width=.98\linewidth]{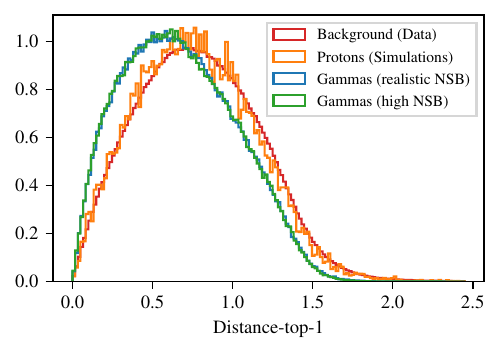}
    \captionof{figure}{
        Distribution of the \distI.
     }
    \label{fig:dist-dist1}
\end{center}
\begin{center}
    
    \includegraphics[width=.98\linewidth]{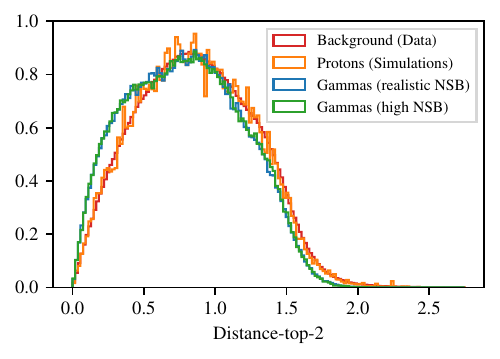}
    \captionof{figure}{
        Distribution of the \distII.
     }
    \label{fig:dist-dist2}
\end{center}
\begin{center}
    
    \includegraphics[width=.98\linewidth]{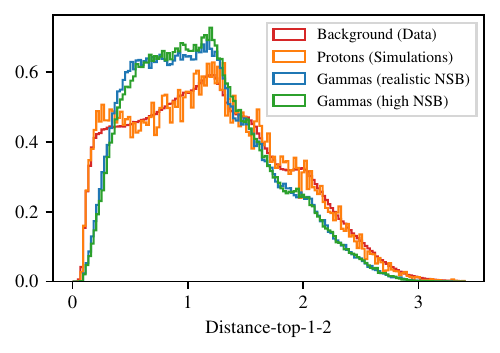}
    \captionof{figure}{
        Distribution of the \distIII.
     }
    \label{fig:dist-dist12}
\end{center}
\begin{center}
    
    \includegraphics[width=.98\linewidth]{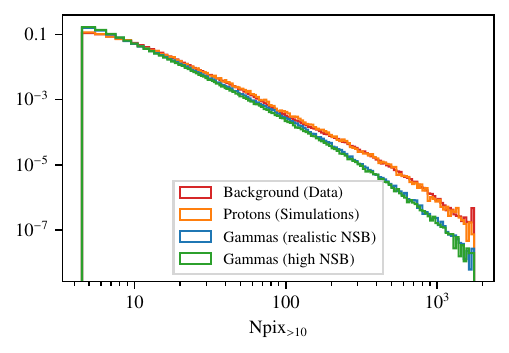}
    \captionof{figure}{
        Distribution of the \npixI.
     }
    \label{fig:dist-n10}
\end{center}
\begin{center}
    
    \includegraphics[width=.98\linewidth]{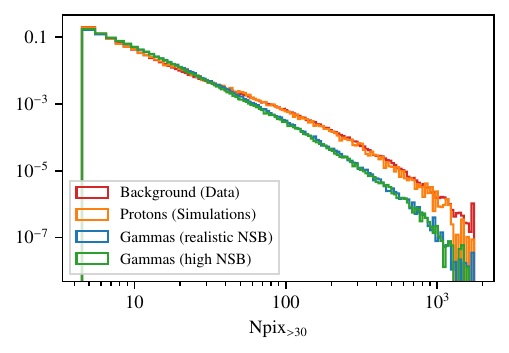}
    \captionof{figure}{
        Distribution of the \npixII.
     }
    \label{fig:dist-n30}
\end{center}

\onecolumn
\clearpage

\section{Variables summary}
In the following we list all variables investigated in this work, segmented in Hillas, sector, continuous, and pixel variables.
\begin{center}
    \centering
    \captionof{table}{Summary table of the variables 
    }
    \label{tab:var_summary}
\begin{tabular}{lccccc}
\hline\hline
Variable & Mirror & Flip & Reco & $\gamma$/h-Sep. & A-stat. \\
\hline
\size &  & x & Yes & x & 2.422 \\
\hlocdist &  & 0.014 & Yes & 0.033 & x \\
\npix &  & 0.018 & Yes & 0.058 & 0.427 \\
\hlength &  & x & Yes & 0.045 & 6.637 \\
\hwidth &  & x & Yes & 0.241 & 58.172 \\
\hskew & $\times (-1)$ & 0.263 & Yes & 0.013 & 8.476 \\
\hkurt &  & x & Yes & 0.008 & 20.770 \\
\hlols &  & 0.008 & Yes & 0.021 & 5.819 \\
\hlogdens &  & x & No & 0.066 & 31.479 \\
\hline
\xO & $-$\xII & x & No & 0.011 & 6.259 \\
\xI &  & x & No & x & x \\
\xII & $-$\xO & x & No & 0.004 & 4.478 \\
\yO & \yII & x & No & x & x \\
\yI &  & x & No & x & x \\
\yII & \yO & x & No & x & x \\
\wO & \wII & 0.028 & No & 0.046 & 32.785 \\
\wI &  & x & No & 0.015 & 20.559 \\
\wII & \wO & 0.026 & No & 0.008 & 29.886 \\
\chargeasym & $\times (-1)$ & x & Yes & x & x \\
\hline
\kinkx &  & x & No & 0.007 & 14.610 \\
\kinkxgof &  & x & No & 0.026 & 34.703 \\
\kinky & $\times (-1)$ & 0.482 & No & 0.011 & 17.520 \\
\kinkygof &  & 0.012 & No & 0.016 & 7.115 \\
\roughI &  & x & No & 0.006 & 13.753 \\
\roughII &  & x & No & 0.012 & 7.860 \\
\pg &  & x & No & 0.186 & 11.150 \\
\pgI & \pgII & 0.013 & No & 0.005 & 9.967 \\
\pgII & \pgI & 0.013 & No & 0.004 & 5.542 \\
\pggof &  & x & No & 0.007 & 39.337 \\
\stda &  & x & Yes & 0.057 & 9.958 \\
\tgx & $\times (-1)$ & x & Yes & x & 809.564 \\
\tgxgof &  & x & No & x & 24190.345 \\
\tgy &  & x & No & x & 2303.658 \\
\tgygof &  & x & No & x & 19008.329 \\
\hline
\ftII &  & x & No & 0.011 & 0.519 \\
\ftI &  & x & No & 0.009 & 1.086 \\
\npixI &  & x & No & 0.018 & 0.574 \\
\npixII &  & x & No & 0.042 & 5.133 \\
\distI &  & x & No & 0.006 & 1.007 \\
\distII &  & x & No & 0.003 & -0.408 \\
\distIII &  & x & No & 0.006 & 2.619 \\
\hline
\end{tabular}
\tablefoot{
          The ``Mirror'' column shows how the variables behaves when being flipped, i.e., no change, change of sign, or change of variable (with sign change).
    Numbers in the columns for the flip and gamma-hadron separation training indicate the relative variable importance, whereas ``x'' denotes that the variable is not used for the respective training. For the direction and energy reconstruction the choice of variables is identical and no importance is estimated from the neural networks. The last column ``A-stat.'' gives the Anderson-Darling test statistics from Sect.~\ref{sec:AD-test}.
      }
\end{center}
    
\end{appendix}

\end{document}